\documentclass[twocolumn,superscriptaddress,showpacs,prd,
aps,amsmath,amssymb,nofootinbib,eqsecnum]{revtex4-1}
\usepackage{hyperref}
\usepackage{graphicx}
\usepackage{bm} 
\usepackage{mathrsfs} 

\usepackage{color}

\newtheorem{lemma}{Lemma}

\newcommand{\be}{\begin{equation}}
\newcommand{\ee}{\end{equation}}
\newcommand{\bea}{\begin{eqnarray}}
\newcommand{\eea}{\end{eqnarray}}
\newcommand{\M}{\mathscr{M}}
\newcommand{\T}{\mathscr{T}}
\newcommand{\Sp}{\mathscr{S}}
\newcommand{\Ci}{\mathscr{C}}
\newcommand{\w}[1]{\bm{#1}}
\newcommand{\vv}[1]{\bm{#1}}
\newcommand{\vw}[1]{\vec{\w{#1}}}
\newcommand{\uu}[1]{\underline{\bm{#1}}}
\newcommand{\vxi}{\vv{\xi}}
\newcommand{\vchi}{\vv{\chi}}
\newcommand{\uxi}{\uu{\xi}}
\newcommand{\uchi}{\uu{\chi}}
\newcommand{\wxis}{\w{\xi}^*}
\newcommand{\wchis}{\w{\chi}^*}
\newcommand{\vxis}{\vec{\wxis}}
\newcommand{\vchis}{\vec{\wchis}}
\newcommand{\R}{\mathbb{R}}
\newcommand{\Lie}[1]{\mathcal{L}_{#1}\,}
\newcommand{\wnab}{\w{\nabla}}
\newcommand{\ph}{\varphi}
\newcommand{\dd}{\bm{\mathrm{d}}}
\newcommand{\eps}{\epsilon}
\newcommand{\weps}{\w{\eps}}
\newcommand{\vep}{\varepsilon}
\newcommand{\diver}{\wnab\!\cdot}
\newcommand{\ups}{\Upsilon}

\begin{document}

\title{Magnetohydrodynamics in stationary and axisymmetric spacetimes: a fully covariant approach}

\author{Eric Gourgoulhon}
\email{eric.gourgoulhon@obspm.fr}
\affiliation{
Laboratoire Univers et Th\'eories, UMR 8102 du CNRS,
Observatoire de Paris, Universit\'e Paris Diderot, F-92190 Meudon, France}
\author{Charalampos Markakis} 
\email{markakis@uwm.edu}
\affiliation{
Department of Physics, University of Wisconsin-Milwaukee, P.O. Box 413,  
Milwaukee, WI 53201}
\author{K\=oji Ury\=u}
\email{uryu@sci.u-ryukyu.ac.jp}
\affiliation{
Department of Physics, University of the Ryukyus, Senbaru, 
Nishihara, Okinawa 903-0213, Japan}
\author{Yoshiharu Eriguchi}
\email{eriguchi@esa.c.u-tokyo.ac.jp}
\affiliation{
Department of Earth Science and Astronomy, Graduate School of Arts and Sciences, University of Tokyo, Komaba, Meguro-ku,
3-8-1, 153-8902 Tokyo, Japan}

\date{3 May 2011}  

\begin{abstract} 
A fully geometrical treatment of general relativistic magnetohydrodynamics (GRMHD) is developed under the hypotheses of perfect conductivity, stationarity and axisymmetry.
The spacetime is not assumed to be circular, which allows for greater generality than the Kerr-type spacetimes usually considered in GRMHD. 
Expressing the electromagnetic field tensor solely in terms of three scalar fields related to the spacetime symmetries, we generalize previously obtained results in various directions. 
In particular, we present the first relativistic version of the Soloviev transfield equation,
subcases of which lead to fully covariant versions of the Grad-Shafranov equation and of the Stokes equation in the hydrodynamical limit. 
We have also derived, as another subcase of the relativistic Soloviev equation, the equation governing magnetohydrodynamical equilibria with purely toroidal magnetic fields in stationary and axisymmetric spacetimes.
\end{abstract} 

\pacs{04.20.-q, 04.40.Dg, 04.40.Nr, 52.30.Cv, 95.30.Qd}

\maketitle
 
\section{Introduction}

General relativistic magnetohydrodynamics (GRMHD) is a rapidly developing field of 
modern astrophysics \cite{Beski10,Font08,Anton_al10}, driven by 
numerous observations of accretion disks around black holes \cite{MosciGDSL09}, jets in active galactic nuclei or microquasars \cite{MeliaSTTC10,Komis10}, gamma ray bursts, hypernovae,
pulsars \cite{Beski10} and strongly magnetized neutrons stars \emph{(magnetars)}. In a first approximation, all these systems are stationary and axisymmetric. 
While GRMHD had been formulated by Lichnerowicz in 1967 \cite{Lichn67}, its 
development for stationary and axisymmetric spacetimes originates in the work of 
Bekenstein and Oron (1978) \cite{BekenO78} (hereafter BO) and Carter (1979) \cite{Carte79}. 
In particular, BO have established two conservation laws associated with the spacetime symmetries, the first of them being a generalization of the Bernoulli theorem to the case of 
a magnetized fluid. Another important step has been the GR generalization of the famous  Grad-Shafranov equation to the Schwarzschild spacetime by Mobarry and Lovelace (1986) \cite{MobarL86} and to the Kerr spacetime by Nitta, Takahashi and Tomimatsu (1991) 
\cite{NittaTT91} and Beskin and Pariev (1993) \cite{BeskiP93}. The extension of the 
Grad-Shafranov equation to the most general stationary and axisymmetric spacetimes has been
performed by Ioka and Sasaki (2003) \cite{IokaS03}, \emph{most general} meaning without
the assumption of \emph{circularity} (also called \emph{orthogonal transitivity}\footnote{Precise definitions are provided below (Sec.~\ref{s:circ}).}), which holds for the Kerr spacetime. 

All the studies mentioned above either (i) involve coordinate-dependent quantities 
or (ii) introduce some extra-structure in spacetime, such as foliations by 2-surfaces, in addition to the canonical structures induced by the two spacetime symmetries (stationarity
and axisymmetry). For instance, two of the fundamental quantities introduced by BO are defined in terms of the components $F_{\alpha\beta}$ and $u^\alpha$ of the electromagnetic tensor and the fluid 4-velocity
by $\omega := - F_{01} / F_{31}$ and $C := F_{31} / (\sqrt{-g} n u^2)$. 
From these expressions, it is not obvious that these quantities are actually
coordinate-independent. 
Another example, related to the feature (ii) mentioned above, 
is the (2+1)+1 decomposition \cite{GourgB93} used
by Ioka and Sasaki \cite{IokaS03,IokaS04} in their study of noncircular spacetimes. 

In this article, we undertake a systematic study of stationary and axisymmetric GRMHD 
relying solely on the spacetime structure induced by the spacetime symmetries. 
To this aim, we make an extensive use of Cartan's exterior calculus, relying on 
the  nature of the electromagnetic field as a 2-form and the well known formulation 
of Maxwell's equations by means of the exterior derivative operator. 
We also employ the possibly less well known formulation of 
hydrodynamics in terms of the fluid vorticity 2-form, originating in the works of Synge \cite{Synge37} and Lichnerowicz \cite{Lichn41}. This enables us to formulate
GRMHD entirely in terms of exterior forms.  
Such an approach is not only elegant and fully covariant, but also  makes easier some
calculations which turn to be tedious in the component approach. 
We pay attention to keeping hypotheses to a strict minimum, which allows us 
to present the results in their most general form, including noncircular spacetimes, 
and to encompass some special cases that had not been considered before, in particular those corresponding to a pure rotational fluid motion (no meridional circulation) or to a purely toroidal magnetic field. 

The plan of the article is as follows. 
In Sec.~\ref{s:stax} we establish the most general form of a stationary and axisymmetric electromagnetic field, independently of any MHD context. In Sec.~\ref{s:perfect_cond}, we introduce the concept of a perfect conductor and in Sec.~\ref{s:MHD} that of a perfect fluid, leading to the MHD-Euler equation. We also derive two Bernoulli-like conservation laws in that section. Section~\ref{s:integrating} is devoted to the integration of the MHD-Euler equation by its reduction to the master transfield equation, a relativistic generalization of the Soloviev transfield equation. 
Various subcases of that equation are examined in Sec.~\ref{s:subcases}, making the link with preceding results in the literature. Finally, Sec.~\ref{s:concl} provides a summary and concluding remarks. 


\section{Stationary and axisymmetric electromagnetic fields} \label{s:stax}

\subsection{Framework and notations}

We consider a spacetime $(\M,\w{g})$, i.e. a four-dimensional real manifold $\M$
endowed with a Lorentzian metric $\w{g}$ of
signature $(- + + +)$. We assume that $\M$ is orientable (see Appendix~\ref{s:diff_geom}), so that we have at our disposal the
Levi-Civita tensor $\weps$ (also called \emph{volume element}) associated with the metric $\w{g}$
[cf. Eq.~(\ref{e:eps_bo})]. Let $\wnab$ be the covariant derivative associated
with $\w{g}$: $\wnab\w{g} = 0$ and $\wnab\weps=0$. 

We shall mostly use an index-free notation, denoting vectors and tensors on $\M$ by boldface symbols. Given a vector $\vv{v}$, we denote by $\uu{v}$ the linear form associated to
$\vv{v}$ by the metric tensor, i.e. the linear form defined by
\be
  \uu{v} := \w{g}(\vv{v},.) .
\ee
Besides, given a linear form $\w{\omega}$, we denote by $\vw{\omega}$ the vector
associated to $\w{\omega}$ by the metric tensor: 
\be \label{e:def_vec_form}
  \w{\omega} =: \w{g}(\vw{\omega},.). 
\ee
In a given basis $(\vv{e}_\alpha)$, where the components of $\w{g}$, $\vv{v}$ and $\w{\omega}$
 are  $g_{\alpha\beta}$, $v^\alpha$ and $\omega_\alpha$ respectively, 
the components of $\uu{v}$ and $\vw{\omega}$ are
$v_\alpha = g_{\alpha\mu} v^\mu$ and $\omega^\alpha = g^{\alpha\mu} \omega_\mu$.

Given a vector $\vv{v}$ and a tensor $\w{T}$ of type $(0,n)$ ($n\geq 1$), i.e. a $n$-linear form (a linear form for $n=1$, a bilinear form for $n=2$, etc.), 
we denote by $\vv{v}\cdot\w{T}$ (resp. $\w{T}\cdot\vv{v}$) the $(n-1)$-linear form
obtained by setting the first (resp. last) argument of $\w{T}$ to $\vv{v}$:
\begin{subequations}
\bea
  \vv{v}\cdot\w{T} & := & \w{T}(\vv{v},.,\ldots,.) \label{e:v_dot_T} \\
  \w{T}\cdot\vv{v} & := & \w{T}(.,\ldots,.,\vv{v}) . 
\eea
\end{subequations}
Thanks to the above conventions, we may write the scalar product of two vectors $\vv{u}$ and $\vv{v}$
as 
\be
  \w{g}(\vv{u},\vv{v}) = \uu{u}\cdot\vv{v} = \vv{u}\cdot\uu{v} . 
\ee
We denote by $\wnab\cdot$ the covariant divergence, with contraction taken on 
the \emph{last} index. For instance, for a tensor field $\w{T}$ of type $(2,0)$, 
$\wnab\cdot\w{T}$ is the vector field defined by
\be 
  \wnab\cdot\w{T} := \nabla_\mu T^{\alpha\mu} \, \vv{e}_\alpha , 
\ee
where $(\vv{e}_\alpha)$ is the vector basis with respect to which the components 
$\nabla_\gamma T^{\alpha\beta}$ of $\wnab\w{T}$ are taken. Note that the convention 
for the divergence does not follow the rule for the contraction with a vector: 
in (\ref{e:v_dot_T}) the contraction is performed on the \emph{first} index.

\subsection{Stationary and axisymmetric spacetimes} \label{s:sta_axi}

We assume that the spacetime $(\M,\w{g})$ possesses
two symmetries: (i) \emph{stationarity}: there exists a group action of $(\R,+)$
on $\M$ whose orbits are timelike curves and which leaves $\w{g}$ invariant;
(ii) \emph{axisymmetry}: there exists a group action of $\mathrm{SO}(2)$ on 
$\M$ whose fixed points form a 2-dimensional submanifold $\Delta\subset\M$ and which 
leaves $\w{g}$ invariant (see e.g. Ref.~\cite{Gourg10} for an extended discussion). 
$\Delta$ is called the \emph{rotation axis}. 
To each parametrization of the one-dimensional Lie groups $(\R,+)$ and
$\mathrm{SO}(2)$, there corresponds
a parametrization of the action orbits; the corresponding tangent vector fields, called
the \emph{generators of the symmetry group}, are denoted $\vxi$ for $(\R,+)$ and $\vchi$ for $\mathrm{SO}(2)$. The invariance of the metric under the actions of $(\R,+)$
and $\mathrm{SO}(2)$ is translated by the vanishing of the Lie derivative of $\w{g}$ along each generator:  
\be  \label{e:sym_metric}
        \Lie{\vxi} \w{g} = 0 \quad\mbox{and}\quad 
        \Lie{\vchi} \w{g} = 0 . 
\ee
The definition of the Lie derivative is recalled in Appendix~\ref{s:Lie_deriv}. 
Thanks to the expression (\ref{e:Lie_der_bilin_nab}) 
and to the identity $\nabla_\mu g_{\alpha\beta} = 0$, Eqs.~(\ref{e:sym_metric}) are equivalent
to the so-called \emph{Killing equations}:
\be \label{e:Killing}
 \nabla_\alpha \xi_\beta + \nabla_\beta \xi_\alpha = 0 \quad\mbox{and}\quad 
         \nabla_\alpha \chi_\beta + \nabla_\beta \chi_\alpha = 0 . 
\ee
The group generators $\vxi$ and $\vchi$ are then called \emph{Killing vectors}. 
For a given group action, a Killing vector is defined up to a constant factor, corresponding to the change of parametrization of the group. 
Regarding the $\mathrm{SO}(2)$ action, we can specify uniquely the Killing vector $\vchi$ by demanding that it corresponds to the standard parametrization of the group
$\mathrm{SO}(2)$, i.e. by selecting the parameter as being the rotation angle 
$\ph\in[0,2\pi[$. For the $(\R,+)$ action, there is a priori no natural scaling of the parameter
$t\in\R$. But if the spacetime is asymptotically flat, we may fix the scaling by demanding that $\vxi$ has the standard normalization at infinity:
\be
        \uxi \cdot \vxi \rightarrow - 1 . 
\ee

For a spacetime that is both stationary and axisymmetric,
Carter \cite{Carte70} has shown that no generality is lost by considering that the stationary and axisymmetric actions commute. 
In other words, the spacetime $(\M,\w{g})$ is invariant under the action  of the \emph{Abelian} group $(\R,+)\times \mathrm{SO}(2)$, and not only under the actions of 
$(\R,+)$ and $\mathrm{SO}(2)$ separately. It is equivalent to say that the 
Killing vectors commute: 
\be \label{e:Kil_commute}
        [\vxi,\vchi] = 0 . 
\ee
Thanks to the property (\ref{e:Kil_commute}), one may introduce 
coordinates $(x^\alpha) = (t,x^1,x^2,\ph)$ on $\M$ such that 
\be \label{e:adapted_coord}
  \vxi = \frac{\partial}{\partial t} \quad\mbox{and}\quad
  \vchi = \frac{\partial}{\partial \ph} . 
\ee
Such coordinates are called \emph{adapted to the spacetime symmetries}.
Within them, 
the metric components are functions of $(x^1,x^2)$ only:
\be
  g_{\alpha\beta} = g_{\alpha\beta}(x^1,x^2) . 
\ee
Adapted coordinate systems are by no means unique: any change of the type
\begin{subequations}
 \label{e:change_coord}
  \bea
    t' & = & t + F_0(x^1,x^2) \label{e:change_coord_t} \\
    {x'}^1 & = & F_1(x^1,x^2) \\
    {x'}^2 & = & F_2(x^1,x^2) \\
    \ph' & = & \ph + F_3(x^1,x^2), \label{e:change_coord_ph} 
  \eea
\end{subequations}
where $F_\alpha(x^1,x^2)$ are well behaved functions, leads to coordinates that are still adapted to the spacetime symmetries.

Using the same notation as Carter in his Les Houches lecture \cite{Carte73},
we introduce the following scalar fields:
\bea
  V & := & - \uxi \cdot \vxi \label{e:def_V} \\
  W & := & \uxi \cdot \vchi \\
  X & := & \uchi \cdot \vchi \\
  \sigma & := & - \det \left[ \begin{array}{ll}
   \uxi \cdot \vxi\  & \uxi \cdot \vchi \\
   \uchi \cdot \vxi\  & \uchi \cdot \vchi
  \end{array} \right] = 
V X + W^2 . \label{e:def_sigma} 
\eea
Since $\vxi$ is assumed to be timelike, we have $V > 0$. Besides, since $\vchi$ is spacelike, $X>0$, except on the rotation axis $\Delta$ where $X=0$. Consequently, 
$\sigma >0$ except on $\Delta$, where $\sigma = 0$. 
For the Minkowski spacetime and using adapted coordinates $(t,r,\theta,\ph)$ of
spherical type, 
\be \label{e:Mink}
   \mbox{\footnotesize Mink.:}\quad V=1,\quad W=0,\quad X = \sigma = r^2\sin^2\theta . 
\ee
We shall also need the Newtonian values of these fields to take non-relativistic limits. In standard isotropic coordinates,
\be \label{e:Newt}
  \mbox{\footnotesize Newt.:}\quad \left\{ 
  \begin{array}{l}
  V = 1 + 2\Phi_{\rm grav}, 
  \qquad W=0 \\
  X = (1 - 2\Phi_{\rm grav}) r^2\sin^2\theta \\
  \sigma = r^2\sin^2\theta ,
  \end{array} \right. 
\ee
where $\Phi_{\rm grav}$ is the Newtonian gravitational potential, 
which obeys  $|\Phi_{\rm grav}| \ll 1$. 
Note that throughout  the article, 
we are using units such that $c=1$. 

\subsection{Orthogonal decomposition of the tangent spaces and circular spacetimes}
\label{s:circ}

The properties of stationarity and axisymmetry define privileged 2-surfaces $\Sp$
in spacetime:
the surfaces of transitivity  of the group action $(\R,+)\times \mathrm{SO}(2)$.
They are spanned by coordinates $(t,\ph)$ of the type (\ref{e:adapted_coord})
and the Killing
vectors $(\vxi,\vchi)$ are everywhere tangent to them. Except on $\Delta$,
$(\vxi,\vchi)$ constitutes a vector basis of the 2-plane $\Pi$ tangent to $\Sp$:
\be  \label{e:def_Pi}
  \Pi = \mathrm{Span}(\vxi,\vchi) . 
\ee
The metric induced by $\w{g}$ in the 2-plane $\Pi$ being non-degenerate ($\Pi$ is a 
timelike plane), the tangent space $\T_x(\M)$ at any point $x\in\M$
can be orthogonally decomposed as the direct sum
\be \label{e:decomp_Pi}
   \T_x(\M) = \Pi \oplus \Pi^\perp , 
\ee
where $\Pi^\perp$ is the (spacelike) 2-plane orthogonal to $\Pi$. 
A vector $\vv{v}\in\T_x(\M)$ is said to be \emph{toroidal} iff $\vv{v}\in\Pi$ with a 
non-vanishing component along $\vchi$ and \emph{poloidal} or
\emph{meridional} iff $\vv{v}\in\Pi^\perp$. 

A question that naturally arises is whether the decomposition (\ref{e:decomp_Pi})
is \emph{integrable}, i.e. whether there exists a family of 2-surfaces such that at every
point $\Pi^\perp$ is tangent to one of these surfaces, in the same way as the $\Pi$ planes
are everywhere tangent to the surfaces of transitivity $\Sp$. 
The spacetimes for which this property holds are called \emph{orthogonally transitive}
or \emph{circular} \cite{Carte69,Carte73}. According to the Frobenius theorem 
(see e.g. Appendix~B of Ref.~\cite{Wald84} or Sec.~7.2 of Ref.~\cite{Strau04}), the necessary and sufficient conditions
for $(\M,\w{g})$ to be circular are
\be \label{e:circ_spacetime}
  \Ci_{\vxi} = 0 \quad\mbox{and}\quad \Ci_{\vchi} = 0 ,
\ee
where $\Ci_{\vxi}$ and $\Ci_{\vchi}$ are the two \emph{twist scalars} defined by
\begin{subequations} \label{e:circul} 
  \bea
   \Ci_{\vxi} & := & \star( \uxi \wedge \uchi \wedge \dd \uxi ) 
  =  \eps^{\mu\nu\rho\sigma} \xi_\mu \chi_\nu \nabla_\rho \xi_\sigma \label{e:def_C_xi} \\
   \Ci_{\vchi}  & := & \star( \uxi \wedge \uchi \wedge \dd \uchi )
  =  \eps^{\mu\nu\rho\sigma} \xi_\mu \chi_\nu \nabla_\rho \chi_\sigma . 
  \eea
\end{subequations}
In these equations, $\dd$ is the exterior derivative, $\wedge$ the exterior product and 
$\star$ the Hodge star; all these operators are defined in Appendix~\ref{s:diff_geom}. 

If $(\M,\w{g})$ is circular, one may choose the adapted coordinates $(t,x^1,x^2,\ph)$
so that $(x^1,x^2)$ span the 2-surfaces orthogonal to the surfaces of transitivity. This leads to the following simplifications in the components of the
metric tensor:
\be \label{e:gab_circular}
  g_{01} = g_{02} = g_{31} = g_{32} = 0 \quad \text{(circular spacetime)}. 
\ee
Examples of circular spacetimes are the Kerr-Newman spacetime (cf. Appendix~\ref{s:Kerr-Newman}) and the spacetime generated by a rotating fluid star with a purely 
poloidal magnetic field \cite{BonazGSM93,BocquBGN95} or a purely toroidal one 
\cite{Oron02,KiuchY08}. In this article, we do not restrict ourselves to the circular case.

\subsection{Stationary and axisymmetric electromagnetic field} \label{s:stax_em}

We consider an electromagnetic field in $\M$, described by the electromagnetic 2-form $\w{F}$, which obeys  Maxwell's equations:
\bea
  & & \dd \w{F} = 0 \label{e:max1} \\
  & & \dd \, \star\!\w{F} = \mu_0 \star\!\uu{j} , \label{e:max2}
\eea 
where $\dd$ is the exterior derivative (cf. Appendix~\ref{s:diff_geom}), 
$\star\w{F}$ is the 2-form Hodge-dual of $\w{F}$ [cf. Eq.~(\ref{e:Hodge_2})]:
\be
  \star\!F_{\alpha\beta} := \frac{1}{2} \eps_{\alpha\beta\mu\nu} F^{\mu\nu} ,
\ee
$\star\uu{j}$ is the 3-form Hodge-dual of the 1-form $\uu{j}$ associated with the electric 4-current $\vv{j}$ [cf. Eq.~(\ref{e:Hodge_1})] : 
\be
  \star\!\uu{j} := \weps(\vv{j}, .,.,.) 
\ee
and $\mu_0$ is the vacuum  permeability. 

We assume that the electromagnetic field is both stationary and axisymmetric. 
This is expressed by the vanishing of the Lie derivatives of the electromagnetic tensor along the symmetry generators $\vxi$ and $\vchi$:
\be \label{e:sym_F}
    \Lie{\vxi} \w{F} = 0 \quad\mbox{and}\quad   \Lie{\vchi} \w{F} = 0 .
\ee
Now, thanks to the Cartan identity (\ref{e:Cartan}) and the Maxwell equation (\ref{e:max1}),
$\Lie{\vxi} \w{F} = \vxi \cdot \dd\w{F} + \dd(\vxi\cdot\w{F}) = \dd(\vxi\cdot\w{F})$. 
Hence Eqs.~(\ref{e:sym_F}) are equivalent to 
\[
        \dd (\vxi\cdot\w{F}) = 0 \quad\mbox{and}\quad  \dd (\vchi\cdot\w{F}) = 0 .
\]
Invoking the Poincar\'e lemma, we conclude that there exist (at least locally) two scalar fields $\Phi$  and $\Psi$ such that
\bea 
    \vxi\cdot\w{F} & = &-  \dd \Phi   , \label{e:def_Phi} \\
     \vchi\cdot\w{F} & =&  - \dd \Psi  . \label{e:def_Psi}
\eea
$\Phi$ and $\Psi$ are defined up to some additive constant and the minus sign is chosen for later convenience.

One very often introduces an electromagnetic 4-potential, i.e. a
 1-form $\w{A}$ such that $\w{F} = \dd\w{A}$. 
Thanks to the identity $\dd\dd\w{A} = 0$ [cf. Eq.~(\ref{e:ext_der_nilpot})], the Maxwell
equation (\ref{e:max1}) is then automatically satisfied. As shown in Appendix~\ref{s:potential}, one may use the gauge freedom on $\w{A}$ to set
\be
\label{e:At_Aphi}
        \Phi = \w{A}\cdot\vxi = A_t  \quad\mbox{and}\quad
        \Psi = \w{A}\cdot\vchi = A_\ph , 
\ee
where the equalities with $A_t$ and $A_\ph$ rely on an adapted coordinate system
$(t,x^1,x^2,\ph)$ [cf. Eq.~(\ref{e:adapted_coord})]. In this article, instead of $\w{A}$ we will use the gauge-independent quantities $\Phi$ and $\Psi$.  

From Eq.~(\ref{e:def_Phi}), $\Lie{\vxi}\Phi = \vxi\cdot \dd\Phi = - \w{F}(\vxi,\vxi) = 0$ and $\Lie{\vchi}\Phi = \vchi\cdot \dd\Phi = - \w{F}(\vxi,\vchi)$.  
Similarly, from Eq.~(\ref{e:def_Psi}), $\Lie{\vchi}\Psi = \vchi\cdot \dd\Psi = - \w{F}(\vchi,\vchi) = 0$ and $\Lie{\vxi}\Psi = \vxi\cdot \dd\Psi =  \w{F}(\vxi,\vchi)$. 
Hence we have
\[
\dd[\w{F}(\vxi,\vchi)] = \dd[\vxi\cdot\dd\Psi] = \Lie{\vxi}\dd\Psi = \Lie{\vxi}(\w{F}\cdot\vchi) = 0 , 
\]
from which we conclude that $\w{F}(\vxi,\vchi)$ is constant over $\M$.
We assume that this constant is zero. In particular this is the case if $\w{F}$ vanishes at some place (e.g. at spatial infinity): 
\be
        \w{F}(\vxi,\vchi) = 0. 
\ee
A consequence of the above property is that the potentials $\Phi$ and $\Psi$ obey both spacetime symmetries: 
\be \label{e:pot_sym}
        \Lie{\vxi}\Phi = \Lie{\vchi}\Phi =0 \quad\mbox{and}\quad
        \Lie{\vxi}\Psi = \Lie{\vchi}\Psi =0 . 
\ee

Apart from $\w{F}(\vxi,\vchi)$, the only non trivial scalar that one can form from 
$\w{F}$, $\vxi$ and $\vchi$ is
\be \label{e:def_I}
        I := \star\!\w{F}(\vxi,\vchi) . 
\ee
We may then assert that the most general form of a stationary and axisymmetric electromagnetic field is 
\be \label{e:F_gal}
        \w{F}  =  \dd \Phi \wedge \wxis + \dd \Psi \wedge \wchis
  + \frac{I}{\sigma} \, \weps(\vxi, \vchi, . ,.) 
\ee
\be \label{e:sF_gal}
        \star\!\w{F}  =  \weps(\vec{\wnab} \Phi, \vxis,.,.)
  + \weps(\vec{\wnab} \Psi, \vchis,.,.)
  - \frac{I}{\sigma} \, \uu{\xi} \wedge \uu{\chi}  , 
\ee
where the 1-forms $(\wxis,\wchis)$ constitute the dual basis of the vector basis
$(\vxi,\vchi)$ of the plane $\Pi$ defined by Eq.~(\ref{e:def_Pi}) and vanish on 
$\Pi$'s orthogonal complement: 
\be
        \wxis  \cdot \vxi = 1, \quad 
        \wxis \cdot \vchi = 0, \quad 
        \wchis\cdot \vxi = 0, \quad 
        \wchis\cdot \vchi = 1,  \label{e:wxi_chi}
\ee
\be \label{e:wxi_chi_perp}
         \forall\vv{v}\in\Pi^\perp,\quad \wxis\cdot \vv{v} = \wchis\cdot\vv{v} = 0 . 
\ee
Conditions (\ref{e:wxi_chi})-(\ref{e:wxi_chi_perp}) define $(\wxis,\wchis)$
uniquely. Indeed it is easy to see that, in terms of the scalars defined 
by (\ref{e:def_V})-(\ref{e:def_sigma}), 
\be \label{e:wxis_uxi_uchi}
        \wxis = \frac{1}{\sigma} (-X\, \uxi + W \, \uchi)\quad\mbox{and}\quad
        \wchis = \frac{1}{\sigma} (W \, \uxi + V \, \uchi) . 
\ee
In terms of coordinates $(t,x^1,x^2,\ph)$ adapted to the spacetime symmetries, 
we may express the 1-forms $\wxis$ and $\wchis$ as
\begin{subequations} \label{e:wxis_dt}
\bea
        \wxis & = & \dd t + \frac{1}{\sigma}(-X \xi_a + W \chi_a) \, \dd x^a \\
        \wchis & = & \dd \ph + \frac{1}{\sigma}(W \xi_a +V \chi_a) \, \dd x^a , 
\eea
\end{subequations}
where the index $a$ ranges from $1$ to $2$. 
In particular, for circular spacetimes, $\xi_a = g_{a0} =0$ and 
$\chi_a = g_{a3} = 0$ [cf. (\ref{e:gab_circular})], so that
\be  \label{e:wxis_circular}
        \wxis = \dd t \quad\mbox{and}\quad
        \wchis = \dd \ph \quad\text{(circular spacetime)}. 
\ee
To demonstrate (\ref{e:F_gal}) let us consider the 2-form 
\[
        \w{G} := \w{F} - \dd \Phi \wedge \wxis - \dd \Psi \wedge \wchis.
\]
It satisfies
\bea
        \w{G}(\vxi,.) & = & \underbrace{\w{F}(\vxi,.)}_{-\dd\Phi}
  - (\underbrace{\vxi\cdot\dd \Phi}_{0}) \wxis
  + (\underbrace{\wxis\cdot \vxi}_{1}) \dd\Phi
  - (\underbrace{\vxi\cdot\dd \Psi}_{0}) \wchis \nonumber \\
  &  & + (\underbrace{\wchis\cdot \vxi}_{0}) \dd\Psi = 0 . \nonumber 
\eea
Similarly, $\w{G}(\vchi,.) = 0$. Hence the 2-form $\w{G}$ vanishes on the plane $\Pi$, 
i.e. the non-trivial action of $\w{G}$ takes place in the plane $\Pi^\perp$. 
Another 2-form that shares the same properties as $\w{G}$ is
\[
        \w{H} :=  \frac{1}{\sigma} \weps(\vxi,\vchi,.,.) . 
\]
Since the vector space of 2-forms in the 2-plane $\Pi^\perp$ is of dimension 1, 
$\left.\w{G}\right|_{\Pi^\perp}$ and $\left.\w{H}\right|_{\Pi^\perp}$ 
must be colinear. Since the $\left.\w{H}\right|_{\Pi^\perp}$  is not vanishing, we conclude that there must exist some coefficient $I$ such that 
$\left.\w{G}\right|_{\Pi^\perp} = I\, \left.\w{H}\right|_{\Pi^\perp}$.
Since both 2-forms vanish on $\Pi$, we 
may extend the equality to $\w{G}$ and $\w{H}$, thanks to the property (\ref{e:decomp_Pi}): 
\[
        \w{G} = I \, \w{H} . 
\]
This proves that $\w{F}$ takes the form (\ref{e:F_gal}). 
Using the properties (\ref{e:star_wedge}), it is then  immediate to show that the Hodge dual of $\w{F}$ is given by 
(\ref{e:sF_gal}). On this form, we verify\footnote{using the fact
that $\weps(\vec{\wnab} \Phi, \vxis,\vxi,\vchi) =0$, thanks to 
relation (\ref{e:wxis_uxi_uchi}) which induces that the vector
$\vxis$ is a linear combination of $\vxi$ and $\vchi$}
that $\star\w{F}(\vxi,\vchi) = I$, i.e. 
that the proportionality coefficient $I$ introduced above is indeed the quantity 
defined by Eq.~(\ref{e:def_I}). This completes the proof of the decomposition 
(\ref{e:F_gal}) of $\w{F}$. 

Equation~(\ref{e:F_gal}) shows that a stationary and axisymmetric electromagnetic field
is entirely described by three scalar fields: 
$\Phi$, $\Psi$ and $I$. 
A concrete example is provided by the 
Kerr-Newman electromagnetic field presented in Appendix~\ref{s:Kerr-Newman}. 
The component expression of (\ref{e:F_gal}) with respect to an adapted coordinate system is given in Appendix~\ref{s:comp}.

\subsection{Maxwell equations}

The first Maxwell equation, Eq.~(\ref{e:max1}), is automatically satisfied by the form (\ref{e:F_gal})
of $\w{F}$, whatever the values of $\Phi$, $\Psi$ and $I$. Indeed, since $\dd \dd \Phi = \dd \dd \Psi=0$ [Eq.~(\ref{e:ext_der_nilpot})], we have, using the Leibniz rule (\ref{e:Leibniz_wedge}) with $p=1$, 
\be \label{e:dF}
  \dd \w{F} = - \dd\Phi \wedge \dd\wxis - \dd\Psi\wedge \dd\wchis + \dd[I\sigma^{-1}\weps(\vxi,\vchi,.,.)] 
\ee
and each of the three terms in the right-hand side vanishes. 
Regarding the first term, we have, via the Cartan identity, 
\[
  \vxi\cdot\dd\wxis = \underbrace{\Lie{\vxi} \wxis}_{0} - \dd(\underbrace{\vxi\cdot\wxis}_{1})
  = 0 . 
\]
Similarly, $\vchi\cdot\dd\wxis = 0$. Hence the 2-form $\dd\wxis$ vanishes on $\Pi$. 
The same thing holds for the 1-form $\dd\Phi$, by virtue of 
Eq.~(\ref{e:pot_sym}). Consequently, the 
3-form $\dd\Phi \wedge \dd\wxis$ vanishes on $\Pi$ and acts only in $\Pi^\perp$. Since 
$\mathrm{dim}\, \Pi^\perp = 2$,  the 3-form $\dd\Phi \wedge \dd\wxis$ necessarily vanishes on $\Pi^\perp$. We thus conclude that $\dd\Phi \wedge \dd\wxis = 0 $ in all space. 
The same property holds for the 3-form $\dd\Psi\wedge \dd\wchis $. 
Finally, regarding the third term in (\ref{e:dF}), let us take its Hodge 
dual and write, using (\ref{e:star_wedge}), 
\[
  \star \dd[I\sigma^{-1}\weps(\vxi,\vchi,.,.)] = - \star\! \dd\star(I\sigma^{-1} \, \uxi\wedge\uchi) . 
\]
Now, the operator $\star\! \dd\star$ is the \emph{codifferential} and can be expressed as the 
\emph{divergence} taken with the $\wnab$ connection: 
$\star\! \dd\star(I\sigma^{-1} \, \uxi\wedge\uchi) = \diver(I\sigma^{-1} \, \uxi\wedge\uchi)$.
Now it is easy to see that $\diver(I\sigma^{-1} \, \uxi\wedge\uchi) = I\sigma^{-1} \, 
\underline{[\vxi,\vchi]} = 0$ by virtue of Eq.~(\ref{e:Kil_commute}). Hence
$\star \dd[I\sigma^{-1}\weps(\vxi,\vchi,.,.)] = 0$, which implies
$\dd[I\sigma^{-1}\weps(\vxi,\vchi,.,.)] = 0$. We conclude that 
(\ref{e:dF}) reduces to $\dd\w{F}=0$, i.e. the first Maxwell equation
(\ref{e:max1}). 

The second Maxwell equation, Eq.~(\ref{e:max2}), gives the electric 4-current $\vv{j}$. 
Let us first fix the first two arguments of each 3-form appearing in Eq.~(\ref{e:max2}) to
$(\vxi,\vchi)$:
\be \label{e:eps_j_xi_chi_prov}
  \mu_0 \weps(\vv{j},\vxi,\vchi,.) = \dd\star\!\w{F}(\vxi,\vchi,.) . 
\ee
Now, by means of the Cartan identity, 
\[
  \dd\star\!\w{F}(\vxi,.,.) = \underbrace{\Lie{\vxi} \star\!\w{F}}_{0}
  - \dd[ \star\! \w{F}(\vxi,.)] =
- \dd[ \star\! \w{F}(\vxi,.)] . 
\]
Hence 
\bea
  \dd\star\!\w{F}(\vxi,\vchi,.) &=& - \vchi \cdot \dd[ \star\! \w{F}(\vxi,.)] \nonumber \\
  & = &- \big\{ \underbrace{\Lie{\vchi}[ \star\! \w{F}(\vxi,.)]}_{0}
  - \dd[  \underbrace{\star\! \w{F}(\vxi,\vchi)]}_{I} \big\} = \dd I . \nonumber  
\eea
Therefore Eq.~(\ref{e:eps_j_xi_chi_prov}) becomes
\be \label{e:eps_j_xi_chi}
  \mu_0 \weps(\vv{j},\vxi,\vchi,.) =  \dd I . 
\ee
We conclude that if the 4-current has some poloidal part, i.e. if $\vv{j}\not\in\Pi$, then
necessarily $I\not = 0$. 

Taking the Hodge dual of (\ref{e:eps_j_xi_chi}) and applying the resulting 3-form to the
couple $(\vxi,\vchi)$ yields
\begin{subequations} \label{e:4current}
\be \label{e:j_gal}
  \vv{j} = (\wxis\cdot\vv{j})\, \vxi + (\wchis\cdot\vv{j})\, \vchi
  - \frac{1}{\mu_0\sigma} \vec{\weps}(\vxi,\vchi,\vec{\wnab}I,.) . 
\ee
There remains to evaluate $\wxis\cdot\vv{j}$ and $\wchis\cdot\vv{j}$. To this aim, let us consider the Maxwell equation (\ref{e:max2}) in its dual form 
$\diver\w{F} = \mu_0 \uu{j}$. Substituting (\ref{e:F_gal}) for $\w{F}$ in it,  expanding and making use of (\ref{e:wxis_uxi_uchi}) and (\ref{e:Killing}) results in 
\bea
 \mu_0 \wxis\cdot\vv{j} & = & \nabla_\mu \left( \frac{X}{\sigma} \nabla^\mu \Phi
  - \frac{W}{\sigma} \nabla^\mu \Psi \right) \nonumber \\
  & & + \frac{I}{\sigma^2} \left( -X \Ci_{\vxi} + W \Ci_{\vchi} \right) \label{e:xis_j} \\
\mu_0 \wchis\cdot\vv{j} & = & -\nabla_\mu \left( \frac{W}{\sigma} \nabla^\mu \Phi
  + \frac{V}{\sigma} \nabla^\mu \Psi \right) \nonumber \\
  & & + \frac{I}{\sigma^2} \left( W \Ci_{\vxi} + V \Ci_{\vchi} \right) , \label{e:chis_j}
\eea
\end{subequations}
where the twist scalars $\Ci_{\vxi}$ and $\Ci_{\vchi}$ are defined by
(\ref{e:circul}). 

At the Newtonian limit, the electric 4-current $\vv{j}$ can be decomposed
into the charge density $\rho_{\rm e}$ and the electric 3-current $\vv{J}$, both measured by the stationary observer, according to 
$\vv{j} = \rho_{\rm e} \, \vxi + \vv{J}$ and $\uxi\cdot\vv{J} = 0$. From 
Eq.~(\ref{e:j_gal}), we get $\rho_{\rm e} = \wxis\cdot\vv{j}$ and 
$\vv{J} = (\wchis\cdot\vv{j})\, \vchi
  - \mu_0\sigma^{-1} \vec{\weps}(\vxi,\vchi,\vec{\wnab}I,.)$.
Choosing spherical coordinates $(t,r,\theta,\ph)$ and using Eq.~(\ref{e:Newt}) as well as Eq.~(\ref{e:weps_dx1dx2}) with
$\sqrt{-g} = r^2\sin\theta$ to express $\vec{\weps}(\vxi,\vchi,\vec{\wnab}I,.)$, we obtain the Newtonian limit of Eqs.~(\ref{e:4current}) as
\begin{subequations} \label{e:current_Newt}
\be
  \mbox{\footnotesize Newt.: \quad}       \mu_0 \rho_{\rm e}  =   \frac{1}{r^2} \partial_r \left( r^2 \partial_r \Phi \right) 
  + \frac{1}{r^2\sin\theta} \partial_\theta \left( \sin\theta \partial_\theta \Phi \right) 
\ee
\be
  \mbox{\footnotesize Newt.:\quad}       \mu_0 \vv{J}  =  \frac{1}{r\sin\theta} \left[ \frac{1}{r} \partial_\theta I \, 
  \vv{e}_{(r)} - \partial_r I \, \vv{e}_{(\theta)} - \Delta^* \Psi \, 
  \vv{e}_{(\ph)} \right] , 
\ee
\end{subequations}
where $(\vv{e}_{(r)}, \vv{e}_{(\theta)}, \vv{e}_{(\ph)})$ is the standard orthonormal basis
associated with spherical coordinates: $\vv{e}_{(r)} := \partial_r$, 
$\vv{e}_{(\theta)} := r^{-1} \partial_\theta$ and
$\vv{e}_{(\ph)} := (r\sin\theta)^{-1} \partial_\ph$, and $\Delta^*$ is the 
second-order differential operator defined by 
\be \label{e:def_Delta_s_Newt}
  \Delta^* \Psi  := \partial^2_r \Psi + \frac{\sin\theta}{r^2}
  \partial_\theta \left( \frac{1}{\sin\theta} \partial_\theta \Psi \right) . 
\ee


\section{Perfect conductor} \label{s:perfect_cond}

\subsection{Definition and first properties}

From now on, we assume that a part $\mathcal{D}\subset\M$ of spacetime is occupied by a 
\emph{perfect conductor}. By this, we mean that $\mathcal{D}$ is covered by a congruence of
timelike worldlines\footnote{later on, we will specify these worldlines to be those of a perfect fluid, but this not necessary for the present discussion} such that the observers associated with each worldline measure a vanishing electric field. This expresses the \emph{infinite conductivity} condition via Ohm's law. Let us recall that the electric field 1-form
$\w{e}$ and the magnetic field vector $\vv{b}$ measured by an observer of 4-velocity $\vv{u}$
are given in terms of $\w{F}$ by 
\be \label{e:def_e_b}
        \vv{e} = \w{F}\cdot \vv{u}
        \qquad\mbox{and}\qquad
        \uu{b} = \vv{u} \cdot\star\!\w{F} . 
\ee
Equivalently, $\w{F}$ is entirely expressible in terms of $\w{e}$, $\vv{b}$ and $\vv{u}$
as
\begin{subequations}
\bea
         \w{F} & = & \uu{u}\wedge \w{e} + \weps(\vv{u},\vv{b},.,.)  \\
        \star \!\w{F} & = & - \uu{u}\wedge \uu{b} + \weps(\vv{u},\vw{e},.,.) . 
\eea
\end{subequations}
The perfect conductor condition is $\vv{e} = 0$. From (\ref{e:def_e_b}), this is equivalent to 
\be  \label{e:perfect_cond}
         \w{F}\cdot \vv{u} = 0 . 
\ee
The electromagnetic field then reduces  to
\begin{subequations}
\bea
         \w{F} & = & \weps(\vv{u},\vv{b},.,.)  \\
        \star \!\w{F} & = & - \uu{u}\wedge \uu{b}  . \label{e:sF_u_b}
\eea
\end{subequations}

Let us decompose the 4-velocity $\vv{u}$ orthogonally with respect to the 2-plane $\Pi$, 
thereby introducing the scalars $\lambda$ and $\Omega$ and the vector $\vv{w}$ : 
\be \label{e:u_expand}
\vv{u} = \lambda (\vxi + \Omega \vchi) + \vv{w}, \qquad
\vv{w} \in \Pi^\perp . 
\ee
The 4-velocity normalization relation $\uu{u}\cdot\vv{u}=-1$ is equivalent to the following 
relation between $\lambda$, $\Omega$ and $\vv{w}$ [cf. Eq.~(\ref{e:def_V})-(\ref{e:def_sigma})]:
\be \label{e:lambda}
\lambda = \sqrt{ \frac{1+\uu{w}\cdot\vv{w}}{V-2W \Omega - X \Omega^2}} . 
\ee
For circular spacetimes and in coordinates $(t,x^1,x^2,\ph)$ adapted to the spacetime symmetries, one has
$w^0 = w^3=0$ [Eq.~(\ref{e:w0_w3_null}) below], 
so that $d\ph/dt = u^3/u^0 = \Omega$, showing that $\Omega$ is the
\emph{angular velocity} of the fluid about the rotation axis. 
On the other side, we shall call $\vv{w}$ the \emph{meridional velocity}.
We shall say that the fluid is in \emph{pure rotational} motion iff $\vv{w} = 0$; the 4-velocity $\vv{u}$ is then a linear combination of the two Killing vectors. 

Cartan's identity (\ref{e:Cartan}), along with Maxwell equation (\ref{e:max1})
and the perfect conductor condition (\ref{e:perfect_cond}) gives
$\Lie{\vv{u}} \w{F} = \vv{u}\cdot \dd \w{F} + \dd(\vv{u}\cdot\w{F}) = 0 + 0$, i.e. 
\be
        \Lie{\vv{u}} \w{F} = 0 . 
\ee
This result is independent of the hypotheses of stationarity and axisymmetry and is the
geometrical expression of
\emph{Alfv\'en's theorem} about magnetic flux freezing. 

From the very definition of $\Phi$ and $\Psi$ [Eqs.~(\ref{e:def_Phi})-(\ref{e:def_Psi})], 
we have $\Lie{\vv{u}}\Phi = \vv{u}\cdot\dd\Phi = - \w{F}(\vxi,\vv{u})$
and $\Lie{\vv{u}}\Psi = - \w{F}(\vchi,\vv{u})$. The
perfect conductor condition (\ref{e:perfect_cond}) gives then immediately
\be \label{e:Lie_u_Phi_Psi}
        \Lie{\vv{u}}\Phi = 0 \qquad\mbox{and}\qquad
        \Lie{\vv{u}}\Psi = 0 . 
\ee
Hence the potentials $\Phi$ and $\Psi$ are preserved along the fluid lines. 
The expansion (\ref{e:u_expand}) of $\vv{u}$ and the symmetry properties (\ref{e:pot_sym})
show that (\ref{e:Lie_u_Phi_Psi}) is actually equivalent to 
\be \label{e:Lie_w_Phi_Psi}
        \Lie{\vv{w}}\Phi = 0 \qquad\mbox{and}\qquad
        \Lie{\vv{w}}\Psi = 0 . 
\ee

Let us express the perfect conductor condition (\ref{e:perfect_cond}) by replacing $\w{F}$
by its expression (\ref{e:F_gal}); we get
\bea
        & & (\underbrace{\w{\xi}^*\cdot\vv{u}}_{\lambda}) \dd\Phi 
  - (\underbrace{\dd\Phi\cdot\vv{u}}_{0}) \w{\xi}^*
  + (\underbrace{\w{\chi}^*\cdot\vv{u}}_{\lambda\Omega}) \dd\Psi \nonumber \\
 & &  \qquad\qquad - (\underbrace{\dd\Psi\cdot\vv{u}}_{0}) \w{\chi}^*
  + \frac{I}{\sigma} 
  \underbrace{\weps(\vxi,\vchi,.,\vv{u})}_{-\weps(\vxi,\vchi,\vv{w},.)} = 0 , \nonumber
\eea
where use has been made of (\ref{e:Lie_w_Phi_Psi}). 
Since $\lambda\not=0$ (otherwise $\vv{u}$ would be null), we obtain
\be \label{e:dPhi_dPsi_Iw}
        \dd\Phi = - \Omega \, \dd\Psi + \frac{I}{\sigma \lambda}
  \weps(\vxi,\vchi,\vv{w},.) . 
\ee

\subsection{Conservation of baryon number and stream function}

If $n$ denotes the baryon number density in the fluid frame, the law of baryon number conservation 
is $\diver(n \vv{u}) = 0$, or equivalently, thanks to the decomposition
(\ref{e:u_expand}) with $\vxi$ and $\vchi$ Killing vectors, 
\be
     \diver(n \vv{w}) = 0.      
\ee
Thanks to the identities (\ref{e:divergence_eps}) and (\ref{e:Hodge_1}), 
we may rewrite the above property as
\be
        \dd( n \star\! \uu{w} ) = 0 . 
\ee
From the Poincar\'e lemma (cf. Appendix~\ref{s:diff_geom}), we conclude that there
exists a 2-form $\w{H}$ such that
\be
        \dd\w{H}  = n \star\! \uu{w} . 
\ee
The above relation is analogous to Maxwell equation (\ref{e:max2}), via the identifications
$\star\w{F} \leftrightarrow \w{H}$ and $\mu_0 \vv{j} \leftrightarrow n \vv{w}$. 
Consequently, the same reasoning that led to Eq.~(\ref{e:eps_j_xi_chi}) results in
\be \label{e:df_w}
  n \, \weps(\vv{w},\vxi,\vchi,.) =  \dd f,  
\ee
where the scalar field $f$ is related to $\w{H}$ by the analogue of Eq.~(\ref{e:def_I}): 
$f := \w{H}(\vxi,\vchi)$. We also have the analogue of Eq.~(\ref{e:j_gal}), with  
$\wxis \cdot\vv{w} = 0$ and $\wchis \cdot\vv{w} = 0$ in addition, since $\vv{w}\in\Pi^\perp$: 
\be \label{e:w_df}
        \vv{w} = - \frac{1}{\sigma n} \, \vec{\weps}(\vxi,\vchi,\vec{\wnab} f,.) . 
\ee
This relation shows that the fluid meridional velocity $\vv{w}$ is entirely described by the scalar field 
$f$; $f$ is called the \emph{stream function} (or \emph{Stokes stream function}). 
From Eq.~(\ref{e:df_w}), we have immediately 
$\vxi\cdot\dd f= 0$ and $\vchi\cdot\dd f = 0$,
which shows that $f$ obeys  the two spacetime symmetries. Moreover, 
a direct consequence of Eq.~(\ref{e:w_df}) is
$\vv{w}\cdot\dd f = 0$. In view of Eq.~(\ref{e:u_expand}), this yields 
\be \label{e:f_conserved}
  \Lie{\vv{u}} f = 0 , 
\ee
i.e. $f$ is
conserved along the fluid lines.

The Newtonian limit of Eq.~(\ref{e:w_df}) is easily taken via Eqs.~(\ref{e:Newt}) and
(\ref{e:weps_dx1dx2}): 
\be \label{e:w_Newt}
  \mbox{\footnotesize Newt.:}\quad 
  \vv{w} = \frac{1}{n r \sin\theta} \left[ \frac{1}{r} \partial_\theta f \, 
  \vv{e}_{(r)} - \partial_r f \, \vv{e}_{(\theta)} \right] , 
\ee
where the notation is the same as in Eq.~(\ref{e:current_Newt}).

Taking  (\ref{e:df_w}) into account, the perfect conductor relation (\ref{e:dPhi_dPsi_Iw}) becomes
\be \label{e:dPhi_dPsi_df}
        \dd\Phi = -\Omega \, \dd\Psi + \frac{I}{\sigma n\lambda} \, \dd f . 
\ee

Thanks to Eq.~(\ref{e:w_df}), 
the condition $\vv{w}\cdot\dd\Phi = 0$ [Eq.~(\ref{e:Lie_w_Phi_Psi})] is equivalent to
\[
        \weps(\vxi,\vchi,\vec{\wnab} f,\vec{\wnab}\Phi) = 0  . 
\]
This relation is satisfied if, and only if, the 1-forms $\dd f$ and $\dd\Phi$ are linearly dependent, 
i.e. if there exist some scalar fields $\alpha$ and $\beta$ not simultaneously vanishing such that  
\be  \label{e:dPhi_df_lin}
        \alpha \, \dd \Phi + \beta \, \dd f = 0 . 
\ee
Similarly the condition $\vv{w}\cdot\dd\Psi = 0$ [Eq.~(\ref{e:Lie_w_Phi_Psi})] leads to the existence of two scalar fields $\alpha'$ and $\beta'$ not simultaneously vanishing such that 
\be \label{e:dPsi_df_lin}
        \alpha' \, \dd \Psi + \beta' \, \dd f = 0 . 
\ee

\subsection{Magnetic field in the fluid frame}

The magnetic field in the fluid frame is obtained by substituting (\ref{e:u_expand})
for $\vv{u}$ and (\ref{e:sF_gal}) for $\star \w{F}$ in Eq.~(\ref{e:def_e_b}). 
We get
\bea
  \vv{b} & = & \frac{\lambda}{\sigma} \Big\{
  \left[ I (W+X \Omega) - \lambda^{-1} \weps(\vxi,\vchi,\vec{\wnab}\Psi,\vv{w})
  \right] \vxi \nonumber \\
   & & \qquad + \left[ I(V-W \Omega) +  \lambda^{-1} \weps(\vxi,\vchi,\vec{\wnab}\Phi,\vv{w})
 \right] \vchi \nonumber \\
  & & \qquad - (W+X \Omega) \, \vec{\weps}(\vxi,\vchi,\vec{\wnab}\Phi,.)\nonumber \\
  & & \qquad -  (V-W \Omega) \, \vec{\weps}(\vxi,\vchi,\vec{\wnab}\Psi,.) \Big\} .
\eea

The above expression is fully general. We may specialize it to a perfect conductor by
expressing $\weps(\vxi,\vchi,\vec{\wnab}\Phi,.)$ via Eqs.~(\ref{e:dPhi_dPsi_df})
and (\ref{e:w_df}) : 
$\weps(\vxi,\vchi,\vec{\wnab}\Phi,.) = - \Omega 
\weps(\vxi,\vchi,\vec{\wnab}\Psi,.)- (I/\lambda)\, \uu{w}$. 
Using (\ref{e:df_w}), we get
\bea
        \vv{b} & = & \frac{\lambda}{\sigma} \Bigg\{
  \left[ I (W+X \Omega) + \frac{1}{\lambda n} \dd f \cdot \vec{\wnab}\Psi
  \right] \vxi \nonumber \\
        & & +  \left[ I \left( V-W \Omega - \frac{\uu{w}\cdot\vv{w}}{\lambda^2}\right)
        + \frac{\Omega}{\lambda n} \dd f \cdot \vec{\wnab}\Psi \right] \vchi \nonumber \\
        & &- (V-2W\Omega - X \Omega^2) \,  \vec{\weps}(\vxi,\vchi,\vec{\wnab}\Psi,.) \nonumber \\
        & & + \frac{I}{\lambda}(W+X \Omega) \, \vv{w} \Bigg\} . \label{e:b_perfect_cond}
\eea
The Newtonian limit of this expression is readily obtained by means of Eqs.~(\ref{e:Newt}) and
(\ref{e:weps_dx1dx2}), and after restoration of $c^{-1}$ factors to cancel velocity terms. 
One gets, with the same notation as in Eq.~(\ref{e:current_Newt}), 
\be \label{e:b_Newt}
 \mbox{\footnotesize Newt.:}\quad       \vv{b} = \frac{1}{r\sin\theta} \left[ \frac{1}{r} \partial_\theta \Psi \, \vv{e}_{(r)}
        - \partial_r \Psi \,  \vv{e}_{(\theta)} + I \, \vv{e}_{(\ph)} \right] . 
\ee

Thanks to the properties $\vxi\cdot\dd\Psi=0$, $\vchi\cdot\dd\Psi=0$ [Eq.~(\ref{e:pot_sym})] 
and $\vv{w}\cdot\dd\Psi=0$ [Eq.~(\ref{e:Lie_w_Phi_Psi})], an immediate consequence of 
expression~(\ref{e:b_perfect_cond}) is 
\be
        \Lie{\vv{b}} \Psi = 0 . 
\ee
Hence, like the fluid lines, the magnetic field lines are contained in constant $\Psi$ hypersurfaces.


\section{Ideal MHD} \label{s:MHD}

\subsection{Perfect fluid model}

From now on, we assume that the fluid is a perfect one, i.e. that its  
energy-momentum tensor is given by
\be
\w{T}^{\rm fl} = (\vep + p) \, \uu{u}\otimes \uu{u} + p \w{g} ,
\ee
where $\vep$ is the proper energy density and $p$ the fluid pressure. 
Moreover, we assume that the fluid is a \emph{simple fluid}, that all the thermodynamic quantities depend only on the entropy density $s$ and on the 
proper baryon number density $n$. In particular, 
\be \label{e:EOS}
  \vep = \vep(s,n) . 
\ee
The above relation is called the \emph{equation of state (EOS)}  of the fluid. 
The \emph{temperature} $T$ and the \emph{baryon chemical potential} $\mu$ are then defined by
\be \label{e:def_T_mu}
  T := \frac{\partial\vep}{\partial s} \qquad\mbox{and}\qquad
  \mu := \frac{\partial \vep}{\partial n} .
\ee
As a consequence of the first law of thermodynamics, $p$ is a function of $(s,n)$ 
entirely determined by (\ref{e:EOS}): 
\be \label{e:p_EOS}
  p = -\vep + T  s + \mu n . 
\ee
Let us introduce the \emph{enthalpy per baryon}, 
\be \label{e:def_h}
  h := \frac{\vep+p}{n} = \mu + T S ,
\ee
where $S$ is the entropy per baryon:
\be \label{e:def_S}
  S := \frac{s}{n} .
\ee
The second equality in (\ref{e:def_h}) is an immediate consequence of (\ref{e:p_EOS}).

\subsection{MHD-Euler equation}

The MHD-Euler equation stems from the conservation law of energy-momentum: 
\be \label{e:cons_enermom}
  \diver (\w{T}^{\rm fl} + \w{T}^{\rm em}) = 0 , 
\ee
where $\w{T}^{\rm em}$ is the energy-momentum tensor of the electromagnetic field. 
As  is well known, 
\be \label{e:divTem}
  \diver \w{T}^{\rm em} = - \w{F}\cdot\vv{j} . 
\ee
On the other side, using the baryon number conservation $\diver(n\, \vv{u})=0$, 
the term $\diver\w{T}^{\rm fl}$ can be decomposed into a part along $\vv{u}$, 
\be \label{e:divTfl_u}
  \vv{u} \cdot \diver\w{T}^{\rm fl} = - n T \, \vv{u}\cdot \dd S
\ee
and a part orthogonal to $\vv{u}$ (see e.g. Ref.~\cite{Gourg06} for details): 
\be \label{e:divTfl_ortho}
  \w{\bot}_{\w{u}} \diver\w{T}^{\rm fl} = 
  n[ \vv{u} \cdot \dd(h \uu{u}) - T \dd S ] . 
\ee
The 2-form $\dd(h \uu{u})$ is called the \emph{vorticity 2-form};  
Eq.~(\ref{e:divTfl_ortho}) obtained first by Synge (1937) \cite{Synge37}
(special relativity and $T=0$), Lichnerowicz (1941) \cite{Lichn41}
(general relativity and $T=0$) and Taub (1959) \cite{Taub59} (general case)
(see also \cite{Carte79}) . 

From the perfect conductor relation (\ref{e:perfect_cond}), we have
$\vv{u}\cdot\w{F}\cdot\vv{j} = \w{F}(\vv{u},\vv{j}) = 0$. Hence Eq.~(\ref{e:divTem})
has no component along $\vv{u}$. Consequently, we deduce from 
Eqs.~(\ref{e:divTem}), (\ref{e:divTfl_u}) and (\ref{e:divTfl_ortho}) that the conservation 
law (\ref{e:cons_enermom}) is equivalent to the system
\bea
  && \Lie{\vv{u}} S = 0 \label{e:S_conserved} \\
  && \vv{u} \cdot \dd(h \uu{u}) - T \dd S = \frac{1}{n} \w{F}\cdot\vv{j} . \label{e:MHD-Euler}
\eea
We shall call Eq.~(\ref{e:MHD-Euler}) the \emph{MHD-Euler} equation. 

\subsection{Conserved quantities along the fluid lines} \label{s:conserv_E_L}

For a pure rotational flow, $\vv{u}$ is a linear combination of the two Killing vectors
[Eq.~(\ref{e:u_expand}) with $\vv{w}=0$] and every scalar field that obeys  the spacetime symmetries is conserved along the fluid lines. This is no longer true for a 
flow with a meridional component ($\vv{w}\not=0$). However, in this case, one can derive two conservation laws of ``Bernoulli'' type, which we investigate here. 

\subsubsection{Derivation}

Contracting the MHD-Euler equation (\ref{e:MHD-Euler}) with the vector 
$\vxi$ leads to
\be \label{e:Euler_xi_prov}
  \vv{u} \cdot \dd (h \uu{u}) \cdot \vxi = \w{F}(\vxi,\vv{j}) / n , 
\ee
where we have used $\vxi\cdot\dd S = 0$ since the entropy per baryon is supposed to 
respect the stationarity symmetry, as well as any fluid
quantity. In particular, $\Lie{\vxi} (h \uu{u}) = 0$ and we deduce from 
Cartan's identity (\ref{e:Cartan}) that
\be \label{e:dhuxi}
  \dd(h \uu{u})\cdot\vxi =   \dd(h\uu{u} \cdot \vxi) . 
\ee
Besides, from the very definition of $\Phi$ [Eq.~(\ref{e:def_Phi})], we have
$\w{F}(\vxi,\vv{j}) = - \vv{j} \cdot \dd \Phi$.
Using expression (\ref{e:j_gal}) for $\vv{j}$, along with the symmetry properties
$\vxi\cdot\dd \Phi = 0$ and $\vchi\cdot\dd \Phi = 0$, leads then to 
\be \label{e:F_xi_j}
  \w{F}(\vxi,\vv{j}) = \frac{1}{\mu_0\sigma} \weps(\vxi,\vchi,\vec{\wnab}I,\vec{\wnab}\Phi) . 
\ee
Thanks to Eqs.~(\ref{e:dhuxi}) and (\ref{e:F_xi_j}), Eq.~(\ref{e:Euler_xi_prov}) becomes
\be \label{e:Euler_xi}
  \Lie{\vv{u}} (h\uu{u} \cdot \vxi) = \frac{1}{\mu_0\sigma n} \weps(\vxi,\vchi,\vec{\wnab}I,\vec{\wnab}\Phi) .
\ee

Since we are considering a flow which is not purely rotational, $\w{w}\not = 0$ and, from
Eq.~(\ref{e:w_df}), $\dd f \not = 0$. Then the linear relation (\ref{e:dPsi_df_lin}) between $\dd \Psi$ and $\dd f$ can be rewritten as 
\be \label{e:dPsi_C_df}
  \dd \Psi = C \, \dd f ,
\ee
where $C$ is some scalar field which is necessarily a function of $f$, as a consequence of the following lemma.
\label{s:sec_lemma}
\begin{lemma} \label{l:functions}
If $z$, $p$ and $y$ are three scalar fields on $\M$ such that
\be \label{e:da_c_db}
  \dd z = p \, \dd y \qquad \mbox{and} \qquad \dd y \not = 0 , 
\ee
then both $z$ and $p$ are functions of $y$, with $p$ being the derivative of
$z$ with respect to $y$:
\be
  z=z(y) \qquad \mbox{and} \qquad p=p(y)=z'(y) . 
\ee
\end{lemma}
\emph{Proof:}
Let us take the exterior derivative of Eq.~(\ref{e:da_c_db}) via Eq.~(\ref{e:Leibniz_wedge}); 
thanks to identities $\dd\dd z =0$ and $\dd\dd y = 0$ [Eq.~(\ref{e:ext_der_nilpot})], we get 
\[
        \dd p \wedge \dd y = 0 . 
\]
If $\dd p\not = 0$, this implies that the hypersurfaces of constant $p$ coincide with the hypersurfaces of constant $y$, from which we deduce that $p$ is a function of $y$. 
If $\dd p=0$, then $p$ is constant and it can still be considered as a function of 
$y$ (constant function). Then we have $\dd z = p(y)\, \dd y$, which shows that $z$ is nothing but a primitive of the function $p(y)$; thus $z=z(y)$ and
$p(y)=z'(y)$, which completes the proof. 

Applying Lemma~\ref{l:functions} to Eq.~(\ref{e:dPsi_C_df}), we obtain
\be \label{e:C_f}
        C = C(f). 
\ee
Since $f$ is preserved along the fluid lines [Eq.~(\ref{e:f_conserved})], we have of course the same property for the function $C$: 
\be \label{e:C_conserved}
        \Lie{\vv{u}} C = 0 . 
\ee

Combining Eqs.~(\ref{e:dPsi_C_df}) and (\ref{e:dPhi_dPsi_df}), we get
\be \label{e:dPhi_D_df}
        \dd\Phi = D \, \dd f, 
\qquad\mbox{with} \quad D := - C \Omega  + \frac{I}{\sigma n\lambda} .
\ee
From Lemma~\ref{l:functions},  we have $D = D(f)$ and 
\be \label{e:Lie_u_D}
        \Lie{\vv{u}} D = 0 . 
\ee

In view of (\ref{e:dPhi_D_df}), Eq.~(\ref{e:Euler_xi}) can be rewritten as
\[
\Lie{\vv{u}} (h\uu{u} \cdot \vxi) =  \frac{D}{\mu_0\sigma n} \weps(\vxi,\vchi,\vec{\wnab}I,\vec{\wnab} f) .
\]
Now, according to Eqs.~(\ref{e:w_df}) and (\ref{e:u_expand}), 
\[
  \weps(\vxi,\vchi,\vec{\wnab}I,\vec{\wnab} f) =  \sigma n\,  \vv{w}\cdot \dd I 
  = \sigma n \, \vv{u}\cdot\dd I = \sigma n \Lie{\vv{u}} I . 
\]
Thus
\[
\Lie{\vv{u}} (h\uu{u} \cdot \vxi) =  \frac{D}{\mu_0} \Lie{\vv{u}} I 
= \Lie{\vv{u}} \left( \frac{DI}{\mu_0} \right) ,  
\]
where the second equality follows from (\ref{e:Lie_u_D}) . 
We conclude that the scalar quantity
\be \label{e:def_E}
  E := - h\uu{u} \cdot \vxi + \frac{DI}{\mu_0}
\ee
is conserved along the fluid lines
\be \label{e:E_conserved}
  \Lie{\vv{u}} E = 0 . 
\ee
Thanks to Eqs.~(\ref{e:u_expand}), we may express $E$ as
\be \label{e:E_express}
  E =   \lambda h (V - W \Omega) + \frac{DI}{\mu_0} . 
\ee
In the limit of a vanishing electromagnetic field ($I=0$ and $C=0$), the conservation law
(\ref{e:E_conserved}) is nothing but the relativistic Bernoulli theorem
(see e.g. Ref.~\cite{Gourg06}).

Repeating the same calculation, but with the Killing vector $\vchi$ instead of $\vxi$, 
we arrive at 
\be \label{e:Euler_chi}
  \Lie{\vv{u}} (h\uu{u} \cdot \vchi) =  \frac{1}{\mu_0\sigma n} \weps(\vxi,\vchi,\vec{\wnab}I,\vec{\wnab}\Psi) ,
\ee
instead of (\ref{e:Euler_xi}).
Substituting Eq.~(\ref{e:dPsi_C_df}) for $\dd \Psi$ and making use of 
Eq.~(\ref{e:w_df}), we get
\[
  \Lie{\vv{u}} (h\uu{u} \cdot \vchi) = \frac{C}{\mu_0} \Lie{\vv{u}} I 
  =  \Lie{\vv{u}} \left( \frac{CI}{\mu_0} \right)  ,
\]
where the second equality follows from (\ref{e:C_conserved}). 
We conclude that the quantity
\be \label{e:def_L}
  L := h\uu{u} \cdot \vchi - \frac{CI}{\mu_0}
\ee
is conserved along the fluid lines:
\be \label{e:L_conserved}
  \Lie{\vv{u}} L = 0 . 
\ee
Thanks to Eq.~(\ref{e:u_expand}), we may express $L$ as
\be \label{e:L_express}
  L = \lambda h (W + X \Omega) - \frac{CI}{\mu_0} .
\ee

The conserved quantities $E$ and $L$ can be considered as functions of $f$:
\be \label{e:E_L_f}
   E = E(f) \qquad \mbox{and} \qquad L = L(f) 
\ee
according to the following lemma: \label{s:sec_lemma2}
\begin{lemma} \label{l:conserved_f}
If $\dd f \not = 0$, 
any scalar field  which obeys to the spacetime symmetries and is preserved along the 
fluid lines
is a function of $f$. 
\end{lemma}
\emph{Proof:} Let $z$ be a scalar field with the above properties. Then
$\Lie{\vxi} z = \vxi\cdot\dd z = 0$ and $\Lie{\vchi} z = \vchi\cdot\dd z = 0$,
which shows that $\vec{\wnab} z \in \Pi^\perp$. 
Moreover, the property 
$\Lie{\vv{u}} z = \vv{w}\cdot\dd z = 0$ with $\vv{w}\not=0$ (since 
$\dd f \not=0$) implies that $\vec{\wnab} z$ lies along the orthogonal direction to $\vv{w}$
in the plane $\Pi^\perp$. The latter being generated by $\vec{\wnab} f$ 
[cf. Eq.~(\ref{e:w_df})], we have that $\dd z = \alpha \, \dd f$ for some coefficient $\alpha$. 
The application of Lemma~\ref{l:functions} then completes the proof.

\subsubsection{Newtonian limits} \label{s:Newt_lim}

To take non-relativistic limits, let us introduce the fluid \emph{mass density} $\rho$ and
\emph{specific enthalpy} $H$ by
\be \label{e:def_rho_H}
  \rho := m_{\rm b} \, n \qquad\mbox{and}\qquad H := \frac{\vep_{\rm int} + p}{\rho} , 
\ee
where $m_{\rm b} = 1.66\times 10^{-27} {\rm\; kg}$ is some mean baryon mass and 
$\vep_{\rm int} := \vep - m_{\rm b} n$ is the fluid internal energy density. 
$H$ is related to $h$ via Eq.~(\ref{e:def_h}): 
\be \label{e:h_H}
  h = m_{\rm b}(1 + H) , 
\ee
with $H\ll 1$ at the non-relativistic limit. 
In view of (\ref{e:Newt}), the expansion of Eq.~(\ref{e:lambda}) leads to 
\be \label{e:lambda_Newt}
  \mbox{\footnotesize Newt.:}\quad \lambda = 1 - \Phi_{\rm grav} + \frac{v^2}{2}  , 
\ee 
where $v^2:= \uu{w} \cdot \vv{w} + \Omega^2 r^2 
\sin^2\theta$.  
Substituting expressions (\ref{e:Newt}), (\ref{e:h_H}) and (\ref{e:lambda_Newt}) into
Eqs.~(\ref{e:E_express}) and (\ref{e:L_express}), we get the
Newtonian limit of the conserved quantities $E$ and $L$: 
\bea \label{e:E_Newt}
  \mbox{\footnotesize Newt.:}& & \quad \frac{E}{m_b} - 1 = H + \Phi_{\rm grav} 
  + \frac{v^2}{2}  + \frac{DI}{\mu_0 m_{\rm b}}   \\
  \mbox{\footnotesize Newt.:}& & \quad \frac{L}{m_b} = \Omega r^2 \sin^2 \theta
  - \frac{CI}{\mu_0 m_{\rm b}}. 
\eea
In the absence of electromagnetic field ($I=0$), we recognize in (\ref{e:E_Newt}) 
the classical Bernoulli integral. 

Besides, if we combine Eqs.~(\ref{e:w_Newt}), (\ref{e:b_Newt}) and (\ref{e:dPsi_C_df}), 
we recover the well known property of colinearity of the poloidal magnetic field and meridional velocity: 
\be \label{e:b_p_colin_w}
  \mbox{\footnotesize Newt.:}\quad \vv{b}_{\rm p} = C n \, \vv{w} , 
\ee
where $\vv{b}_{\rm p}$ is the part of $\vv{b}$ along $\vv{e}_{(r)}$ and
$\vv{e}_{(\theta)}$ in Eq.~(\ref{e:b_Newt}).

\subsubsection{Comparison with BO}

The conservation laws (\ref{e:E_conserved}) and (\ref{e:L_conserved}) have been first
established by BO \cite{BekenO78}.
They have expressed $E$ and $L$ in terms of the magnetic field $\vv{b}$ in the fluid
frame, but it can be shown that their expressions are equivalent to
(\ref{e:E_express}) and (\ref{e:L_express}). 
Actually, our derivation is slightly more general. Indeed, the BO 
expressions for $E$ and $L$ are\footnote{Eqs.~(\ref{e:E_BO}) and (\ref{e:L_BO}) are
respectively Eqs.~(92) and (93) of Ref.~\cite{BekenO78}; to show that they are equivalent to
Eqs.~(\ref{e:E_express}) and (\ref{e:L_express}), the starting point is to use the definition
 (\ref{e:def_I}) of $I$ along with the perfect conductor expression (\ref{e:sF_u_b}) 
of $\star \!\w{F}$ to write $I = (\uu{u}\cdot\vchi)(\uu{b}\cdot\vxi) - (\uu{u}\cdot\vxi)
  (\uu{b}\cdot\vchi)$.}
\be \label{e:E_BO}
 E = -  \left( h +  \frac{\uu{b}\cdot\vv{b}}{\mu_0 n} \right) 
                \uu{u}\cdot\vxi - \frac{C}{\mu_0}
  \left[ \uu{u}\cdot(\vxi+\omega \vchi)\right] (\uu{b}\cdot\vxi) ,   
\ee
\be \label{e:L_BO}
  L = \left( h +  \frac{\uu{b}\cdot\vv{b}}{\mu_0 n} \right) 
                \uu{u}\cdot\vchi + \frac{C}{\mu_0}
 \left[ \uu{u}\cdot(\vxi+\omega \vchi) \right] (\uu{b}\cdot\vchi) , 
\ee
where $\omega$ is defined by BO in terms of the components of 
the electromagnetic field tensor as\footnote{$\omega$ is denoted $-A$ by BO.}
\be
  \omega := - \frac{F_{01}}{F_{31}} = - \frac{F_{02}}{F_{32}} . 
\ee
Using Eqs.~(\ref{e:F_0a})-(\ref{e:F_3a}), we see that, within our notations, $\omega$ 
is the proportionality factor between the gradients of $\Phi$ and $\Psi$: 
\be \label{e:dPhi_omega_dPsi}
  \dd \Phi = - \omega \, \dd \Psi . 
\ee
Combining Eqs.~(\ref{e:dPhi_dPsi_df}) and (\ref{e:dPsi_C_df}), we get an expression of 
$\omega$ in terms of previously introduced quantities:
\be \label{e:omega_Omega_I}
  \omega = \Omega - \frac{I}{C \sigma n \lambda} = - \frac{D}{C} . 
\ee
On Eq.~(\ref{e:dPhi_omega_dPsi}) we see the slight shortcoming of BO expressions
for $E$ and $L$: if the electromagnetic field is such that $\dd \Psi = 0$ 
while $\dd\Phi\not=0$ (purely toroidal magnetic field, cf. Sec.~\ref{s:toroidal}), then 
$\omega$ is ill defined: $\omega \rightarrow\infty$. This corresponds to 
$F_{31}=F_{32}=0$ or $C=0$ [cf. Eq.~(\ref{e:dPsi_C_df})]. 
In contrast, our expressions (\ref{e:E_express}) and (\ref{e:L_express}) for $E$ and $L$, and the derivation of their constancy along the streamlines, are valid even in the 
special case $\dd\Psi = 0$. Note however that BO formulas (\ref{e:E_BO})-(\ref{e:L_BO})
give finite expressions when $\omega\rightarrow\infty$ ($\iff  C \rightarrow 0$), since  
Eq.~(\ref{e:omega_Omega_I}) shows that
\[
  C \omega = C \Omega - \frac{I}{\sigma n \lambda} \longrightarrow 
  - \frac{I}{ \sigma n \lambda} \qquad\mbox{when}\quad C \rightarrow 0 . 
\]


\section{Integrating the MHD-Euler equation} \label{s:integrating}

\subsection{Explicit form of the MHD-Euler equation}

Let us first evaluate the 1-form $\vv{u} \cdot \dd(h \uu{u})$ that appears in the 
left-hand side of the MHD-Euler equation (\ref{e:MHD-Euler}), by means of the
decomposition (\ref{e:u_expand}) of $\vv{u}$. We first decompose the
1-form $\vv{u} \cdot \dd(h \uu{u})$ orthogonally with respect to the plane $\Pi$
by writing
\[
  \vv{u} \cdot \dd(h \uu{u}) = \w{Z} + \alpha \, \wxis + \beta \, \wchis , 
\]
where $\w{Z}$ is a 1-form that vanishes in $\Pi$ and the coefficients $\alpha$
and $\beta$ are determined via the properties~(\ref{e:wxi_chi}): 
$\alpha = \vv{u} \cdot \dd(h \uu{u}) \cdot \vxi$ 
and $\beta = \vv{u} \cdot \dd(h \uu{u}) \cdot \vchi$. 
Using the Cartan identity (\ref{e:Cartan}), we get
\bea
 \alpha &= &- [\vxi \cdot \dd(h \uu{u}) ]\cdot \vv{u} = - \big[ 
  \underbrace{\Lie{\vxi}(h \uu{u})}_{0} - \dd(h \uu{u}\cdot\vxi) \big] \cdot \vv{u} \nonumber \\
  &= &\vv{u}\cdot \dd(h \uu{u}\cdot\vxi)  = \vv{w}\cdot \dd(h \uu{u}\cdot\vxi) . \nonumber
\eea
Similarly $\beta = \vv{w}\cdot \dd(h \uu{u}\cdot\vchi)$. Hence
\be \label{e:udhu_Z}
  \vv{u} \cdot \dd(h \uu{u}) = \w{Z} + \left[ \vv{w}\cdot \dd(h \uu{u}\cdot\vxi) \right]  \wxis
  + \left[ \vv{w}\cdot \dd(h \uu{u}\cdot\vchi) \right]  \wchis .
\ee

Besides, from the decomposition (\ref{e:u_expand}) of $\vv{u}$, we have
\be \label{e:udhu_udr}
  \vv{u} \cdot \dd(h \uu{u}) =  \vv{u} \cdot \dd\w{r} + \vv{u} \cdot \dd(h \uu{w})  ,
\ee
where we have introduced the 1-form
\be \label{e:def_r}
  \w{r} := \lambda h ( \uxi  + \Omega \uchi) . 
\ee
We have, using the Cartan identity, 
\bea
  \vv{u} \cdot \dd\w{r}  & = & \lambda \vxi\cdot\dd\w{r} 
  + \lambda\Omega \vchi\cdot\dd\w{r} + \vv{w} \cdot \dd\w{r} \nonumber \\
  & = & \lambda \big[ \underbrace{\Lie{\vxi} \w{r}}_{0} - \dd(\w{r}\cdot\vxi) \big]
  + \lambda\Omega \big[ \underbrace{\Lie{\vchi} \w{r}}_{0} - \dd(\w{r}\cdot\vchi) \big]
  \nonumber \\
  & & + \vv{w} \cdot \dd\w{r} \nonumber \\
  & = & - \lambda \, \dd(h\uu{u}\cdot\vxi)
  - \lambda \Omega \, \dd(h\uu{u}\cdot\vchi) +  \vv{w} \cdot \dd\w{r} \label{e:u_dr} .
\eea
Invoking  the Cartan identity again, 
\[
  \vxi \cdot \dd (h \uu{w}) = \underbrace{\Lie{\vxi} (h\uu{w})}_{0} 
  - \dd( h\underbrace{\uu{w}\cdot\vxi}_{0}) = 0 . 
\]
Similarly, $\vchi \cdot \dd (h \uu{w}) =0$. This shows that the 2-form $\dd (h \uu{w})$ acts only
the 2-plane $\Pi^\perp$. By the same reasoning as for the 2-form $\w{G}$ in 
Sec.~\ref{s:stax_em}, we deduce that $\dd (h \uu{w})$ must be proportional to the
2-form $\weps(\vxi,\vchi,.,.)$:
\be \label{e:dhw_q}
  \dd (h \uu{w}) = q\, \weps(\vxi,\vchi,.,.) .
\ee 
The coefficient $q$ is determined by the Hodge duality:
\be \label{e:q_Delta_f}
  q = \frac{1}{\sigma} \epsilon^{\mu\nu\rho\lambda} \xi_\mu \chi_\nu
  \, \nabla_\rho (h w_\lambda) 
   = - \nabla_\mu \left( \frac{h}{\sigma\, n} \nabla^\mu f \right) ,  
\ee
where the second equality results from Eq.~(\ref{e:w_df}).

Collecting (\ref{e:u_dr}) and (\ref{e:dhw_q}), we rewrite
(\ref{e:udhu_udr}) as 
\bea
  \vv{u} \cdot \dd(h \uu{u}) & = & - \lambda \, \dd(h\uu{u}\cdot\vxi)
  - \lambda \Omega \, \dd(h\uu{u}\cdot\vchi) +  \vv{w} \cdot \dd\w{r} \nonumber \\
  & & + \frac{q}{n} \, \dd f , \label{e:udhu_wdr_q}
\eea
where we have used the property (\ref{e:df_w}): 
$\weps(\vxi,\vchi,\vv{u},.) = \weps(\vxi,\vchi,\vv{w},.) = n^{-1} \dd f$.

Let us employ (\ref{e:udhu_wdr_q}) to evaluate the 1-form $\w{Z}$ acting in the plane $\Pi^\perp$. 
Given a generic vector $\vv{v}\in\Pi^\perp$, we have
\bea
  \w{Z}\cdot\vv{v} & =  & \vv{u} \cdot \dd(h \uu{u}) \cdot \vv{v} \nonumber \\
   & = & - \lambda \, \vv{v}\cdot\dd(h\uu{u}\cdot\vxi)
  - \lambda \Omega \, \vv{v}\cdot \dd(h\uu{u}\cdot\vchi) +  
  \dd\w{r}(\vv{w},\vv{v}) \nonumber \\
   & & + \frac{q}{n}\, \vv{v}\cdot\dd f . \label{e:Z_v}
\eea
There remains to evaluate $\dd\w{r}(\vv{w},\vv{v})$; from (\ref{e:def_r}), we have
\[
  \dd\w{r}(\vv{w},\vv{v}) = \lambda h \left[ \dd\uxi(\vv{w},\vv{v}) 
   + \Omega \, \dd\uchi(\vv{w},\vv{v}) \right] , 
\]
where we have used the property $(\dd\Omega\wedge \uchi)(\vv{w},\vv{v}) = 0$,
resulting from $\uchi\cdot\vv{w}=0$ and $\uchi\cdot\vv{v}=0$. 
By a straightforward calculation\footnote{One may employ formula~(\ref{e:ext_prod_1_2}) to
express $\uchi\wedge\dd\uxi$ and formula~(\ref{e:ext_prod}) with $p=1$ and $q=3$
to compute $\uxi\wedge( \uchi\wedge\dd\uxi)$ on the quadruplet
$(\vxi,\vchi,\vv{w},\vv{v})$.} one can show the identity
\[
  (\uxi\wedge\uchi\wedge\dd\uxi)(\vxi,\vchi,\vv{w},\vv{v}) = - \sigma \, 
  \dd\uxi(\vv{w},\vv{v}) . 
\]
Now, the Hodge dual of relation (\ref{e:def_C_xi}) gives
\[
  \uxi\wedge\uchi\wedge\dd\uxi = - \Ci_{\vxi}\, \weps. 
\]
Hence
\[
  \dd\uxi(\vv{w},\vv{v}) = \frac{\Ci_{\vxi}}{\sigma}
  \, \weps(\vxi,\vchi,\vv{w},\vv{v}) = \frac{\Ci_{\vxi}}{\sigma n} \vv{v}\cdot \dd f , 
\]
where the second equality results from Eq.~(\ref{e:df_w}). 
Using the similar relation for $\dd\uchi(\vv{w},\vv{v})$, we arrive at
\[
  \dd\w{r}(\vv{w},\vv{v}) = \frac{\lambda h}{\sigma n} ( \Ci_{\vxi}
   + \Omega \Ci_{\vchi} )  \,  \vv{v}\cdot \dd f  .
\]

\begin{widetext}

Substituting in Eq.~(\ref{e:Z_v}), we get 
\[
  \w{Z} =  -\lambda \, \dd(h\uu{u}\cdot\vxi)
  - \lambda \Omega \, \dd(h\uu{u}\cdot\vchi) + \frac{1}{n} \left[ q + \frac{\lambda h}{\sigma} \left( \Ci_{\vxi}
   + \Omega \Ci_{\vchi} \right) \right] \, \dd f  . 
\]
Finally, Eq.~(\ref{e:udhu_Z}) becomes
\be
  \vv{u} \cdot \dd(h \uu{u}) = \left[ \vv{w}\cdot \dd(h \uu{u}\cdot\vxi) \right]  \wxis
  + \left[ \vv{w}\cdot \dd(h \uu{u}\cdot\vchi) \right]  \wchis   + \frac{1}{n} \left[ q + \frac{\lambda h}{\sigma} \left( \Ci_{\vxi}
   + \Omega \Ci_{\vchi} \right) \right] \, \dd f  - \lambda \, \dd(h\uu{u}\cdot\vxi)
  - \lambda \Omega \, \dd(h\uu{u}\cdot\vchi) . \label{e:udhu}
\ee

Let us now evaluate the Lorentz force term on the right-hand side of the MHD-Euler equation (\ref{e:MHD-Euler}). Given the generic form (\ref{e:F_gal}) of $\w{F}$, we have
\[
        \w{F}\cdot\vv{j}  =  (\wxis\cdot\vv{j})\, \dd \Phi - (\vv{j}\cdot\dd\Phi) \, \wxis  
        + (\wchis\cdot\vv{j})\, \dd \Psi - (\vv{j}\cdot\dd\Psi) \, \wchis  
  + \frac{I}{\sigma} \, \weps(\vxi, \vchi, . ,\vv{j}) .  
\]
Now, from Eqs.~(\ref{e:j_gal}) and (\ref{e:pot_sym}), 
$\vv{j}\cdot\dd\Phi = 
  - \weps(\vxi,\vchi,\vec{\wnab}I,\vec{\wnab}\Phi) / (\mu_0\sigma)$, 
and $\vv{j}\cdot\dd\Psi = 
  - \weps(\vxi,\vchi,\vec{\wnab}I,\vec{\wnab}\Psi) / (\mu_0\sigma)$. Besides, from 
Eq.~(\ref{e:eps_j_xi_chi}), 
$\weps(\vxi, \vchi, . ,\vv{j}) = - \mu_0^{-1} \, \dd I$. 
Hence 
\be \label{e:Lorentz_force}
        \w{F}\cdot\vv{j}  = \frac{1}{\mu_0\sigma} \weps(\vxi,\vchi,\vec{\wnab}I,\vec{\wnab}\Phi) \, \wxis 
        + \frac{1}{\mu_0\sigma} \weps(\vxi,\vchi,\vec{\wnab}I,\vec{\wnab}\Psi) \, \wchis 
        + (\wxis\cdot\vv{j})\, \dd \Phi + (\wchis\cdot\vv{j})\, \dd \Psi
        - \frac{I}{\mu_0\sigma} \, \dd I . 
\ee

In view of Eqs.~(\ref{e:udhu}) and (\ref{e:Lorentz_force}), the MHD-Euler equation (\ref{e:MHD-Euler}) becomes
\bea
        & & \left[ \vv{w}\cdot \dd(h \uu{u}\cdot\vxi) - \frac{1}{\mu_0\sigma n} \weps(\vxi,\vchi,\vec{\wnab}I,\vec{\wnab}\Phi) \right] \, \wxis 
        + \left[\vv{w}\cdot \dd(h \uu{u}\cdot\vchi) - \frac{1}{\mu_0\sigma n} \weps(\vxi,\vchi,\vec{\wnab}I,\vec{\wnab}\Psi)\right]  \, \wchis 
        + \frac{I}{\mu_0\sigma n} \, \dd I 
\nonumber \\
        & & - \lambda \, \dd(h\uu{u}\cdot\vxi)
  - \lambda \Omega \, \dd(h\uu{u}\cdot\vchi) + \frac{1}{n} \left[ q + \frac{\lambda h}{\sigma} \left( \Ci_{\vxi}
   + \Omega \Ci_{\vchi} \right) \right] \, \dd f 
   - \frac{\wxis\cdot\vv{j}}{n} \, \dd \Phi
   - \frac{\wchis\cdot\vv{j}}{n} \, \dd \Psi
    - T \, \dd S  = 0 . \label{e:MHD-Euler-expl}
\eea
This equation expresses the vanishing of a 1-form. The parts along $\wxis$ and $\wchis$ 
vanish identically in the 2-plane $\Pi^\perp$ [cf. Eq.~(\ref{e:wxi_chi_perp})]. On the contrary,
all the remaining parts, being proportional to gradient of symmetric scalar fields, 
vanish identically in the 2-plane $\Pi = \mathrm{Span}(\vxi,\vchi)$. Each tangent space to $\M$ being the direct sum 
of $\Pi$ and $\Pi^\perp$ [Eq.~(\ref{e:decomp_Pi})] and $(\wxis,\wchis)$ being a basis of the dual space to 
$\Pi$, we deduce that Eq.~(\ref{e:MHD-Euler-expl}) is equivalent to the system of three equations:
\begin{subequations} 
\label{e:syst_MHD_Euler}
\bea
        & & \vv{w}\cdot \dd(h \uu{u}\cdot\vxi) - \frac{1}{\mu_0\sigma n} \weps(\vxi,\vchi,\vec{\wnab}I,\vec{\wnab}\Phi) = 0   \label{e:syst_MHD_Euler-1} \\
        & & \vv{w}\cdot \dd(h \uu{u}\cdot\vchi) - \frac{1}{\mu_0\sigma n} \weps(\vxi,\vchi,\vec{\wnab}I,\vec{\wnab}\Psi) = 0 \label{e:syst_MHD_Euler-2} \\
        & & \lambda \, \dd(h\uu{u}\cdot\vxi)
  +\lambda \Omega \, \dd(h\uu{u}\cdot\vchi) - \frac{1}{n} \left[ q + \frac{\lambda h}{\sigma} \left( \Ci_{\vxi}
   + \Omega \Ci_{\vchi} \right) \right] \, \dd f 
   + \frac{\wxis\cdot\vv{j}}{n} \, \dd \Phi
   + \frac{\wchis\cdot\vv{j}}{n} \, \dd \Psi
   - \frac{I}{\mu_0\sigma n} \, \dd I  \nonumber \\
   & & \ \qquad + T \, \dd S = 0 . \label{e:syst_MHD_Euler-3}
\eea
\end{subequations} 

\end{widetext}

\subsection{Introducing the master potential}

In view of relations~(\ref{e:dPhi_df_lin}) and (\ref{e:dPsi_df_lin}), the 
three linear forms $\dd \Phi$,  $\dd \Psi$ and $\dd f$ are colinear to each other. 
If one of the fields $\Phi$,  $\Psi$ or $f$ is such that it gradient is non-vanishing, then
by virtue of Lemma~\ref{l:functions} (cf. Sec.~\ref{s:sec_lemma}), the two other fields can be
considered as function of it. 
The standard approach in GRMHD is to privilege the field $\Psi$. However this leads to degenerate
equations when $\dd\Psi=0$, which corresponds to purely toroidal magnetic fields or to the hydrodynamical limit (vanishing electromagnetic field). The same problem occurs if one selects $\Phi$ or $f$ instead of $\Psi$ (for instance selecting $f$ leads to degenerate equations in the case of a pure rotational flow). To be fully general, we adopt instead an approach introduced in 
non-relativistic MHD by Tkalich \cite{Tkali59,Tkali62} and Soloviev \cite{Solov67}, namely we consider
a fourth field $\ups$ such that (i) $\ups$ obeys  both spacetime symmetries, (ii)
$\dd \ups$ is never vanishing and (iii) there exists three scalar fields $\alpha$, 
$\beta$ and $\gamma$ such that 
\be \label{e:dall_dups}
  \dd\Phi = \alpha \, \dd\ups, \quad
  \dd\Psi = \beta \, \dd\ups, \quad \dd f = \gamma\, \dd \ups . 
\ee
The existence of $\ups$ is guaranteed by the colinearity properties (\ref{e:dPhi_df_lin}) and (\ref{e:dPsi_df_lin}). Of course, $\ups$ is far from being unique. The special cases mentioned above correspond to $\beta = 0$ or $\gamma = 0$,
with $\dd\ups$ remaining non-vanishing. 
According to Lemma~\ref{l:functions}, $\Phi$, $\Psi$ and $f$ are necessarily functions
of $\ups$, with $\alpha$, $\beta$ and $\gamma$ being their derivatives: 
\bea
  & & \Phi = \Phi(\ups),\quad \Psi = \Psi(\ups), \quad f = f(\ups), 
  \label{e:all_funct_ups} \\
  & & \alpha = \Phi'(\ups), \quad \beta = \Psi'(\ups), \quad \gamma = f'(\ups) . 
\eea
We call $\ups$ the \emph{master potential}. 
Using this  fourth potential allows to treat all cases with finite quantities,
whereas sticking to the three potentials $\Phi$, $\Psi$ and $f$
leads to infinite quantities in the degenerate cases mentioned above. 
In this respect there is some analogy with the use of homogeneous 
coordinates in projective geometry: using only two coordinates $(x,y)$ in the projective
plane $\mathbb{RP}^2$ leads to infinite values for the ``points at infinity'', whereas
adding a third coordinate, forming the so-called homogeneous coordinates $(x,y,z)$, 
fix this, at the price of some redundancy: $(x,y,z)$ and $(\lambda x,\lambda y, \lambda z)$ 
with $\lambda\not=0$ describe the same point, as $\ups$ and $\lambda\ups$ correspond to the 
same configuration. 

The master potential is conserved along any given fluid line. Indeed, if $\dd f\not =0$,
then $\gamma\not=0$ and $\Lie{\vv{u}} \ups = \vv{u} \cdot\dd \ups = \gamma^{-1} \vv{u}\cdot\dd f = 0$ by 
virtue of Eq.~(\ref{e:f_conserved}). If $\dd f = 0$, then $\vv{u}$ is a linear combination of
the Killing vectors $\vxi$ and $\vchi$ and $\Lie{\vv{u}} \ups = 0$ holds according to the hypothesis (i) above. We conclude that in all cases
\be \label{e:ups_conserved}
  \Lie{\vv{u}} \ups = 0 .  
\ee
Besides, in view of (\ref{e:dall_dups}), 
the perfect conductivity relation (\ref{e:dPhi_dPsi_df}) is equivalent to 
\be \label{e:perfect_alpha_beta}
  \alpha + \Omega \beta =  \frac{\gamma I}{\sigma n\lambda} . 
\ee

Let us proceed by rewriting MHD-Euler system~(\ref{e:syst_MHD_Euler}) in terms of
$\ups$. 
Thanks to Eq.~(\ref{e:w_df}), the term $\vv{w}\cdot \dd(h \uu{u}\cdot\vxi)$ in 
Eq.~(\ref{e:syst_MHD_Euler-1}) can be rewritten as
$-(\sigma n)^{-1} \weps(\vxi,\vchi,\vec{\wnab}f, \vec{\wnab}(h \uu{u}\cdot\vxi))$.
Using (\ref{e:dall_dups}), Eq.~(\ref{e:syst_MHD_Euler-1}) is thus equivalent to 
\[
  \weps\left( \vxi,\; \vchi,\; \vec{\wnab}\ups,\; 
  - \gamma \vec{\wnab} (h \uu{u}\cdot\vxi) + \frac{\alpha}{\mu_0} \vec{\wnab} I 
  \right) = 0 . 
\]
Since $\gamma = \gamma(\ups)$ and $\alpha = \alpha(\ups)$, the Leibniz rule and the
alternate character of $\weps$ allow us to write this relation as
\[
  \weps\left( \vxi,\; \vchi,\; \vec{\wnab}\ups,\; 
  \vec{\wnab} \left(- \gamma h \uu{u}\cdot\vxi + \frac{\alpha I}{\mu_0}  \right)
  \right) = 0 . 
\]
This implies that the 1-forms $\dd \ups$ and 
$\dd (-\gamma h \uu{u}\cdot\vxi + \alpha I/\mu_0 )$  are colinear.
Since $\dd \ups\not=0$, we conclude that there exists a scalar field $a$ such that
\[
  \dd \left(- \gamma h \uu{u}\cdot\vxi + \frac{\alpha I}{\mu_0}  \right)
  = a \, \dd\ups. 
\]
Invoking again Lemma~\ref{l:functions}, we conclude that 
$(-\gamma h \uu{u}\cdot\vxi + \alpha I/\mu_0 )$ must be a function of $\ups$,
$\Sigma(\ups)$ say. Expressing $\uu{u}\cdot\vxi$ via Eqs.~(\ref{e:u_expand}) and (\ref{e:def_V})-(\ref{e:def_sigma}), we get 
\be \label{e:Sigma_def}
  \Sigma(\ups) = -\gamma h \uu{u}\cdot\vxi + \frac{\alpha I}{\mu_0} =  \gamma \lambda h (V - W\Omega) + \frac{\alpha I}{\mu_0} .  
\ee

Applying a similar argument to the second equation of the MHD-Euler system  (\ref{e:syst_MHD_Euler}) leads to the existence of a function
$\Lambda(\ups)$ such that
\be \label{e:Lambda_def}
  \Lambda(\ups) = \gamma h \uu{u}\cdot\vchi - \frac{\beta I}{\mu_0} = \gamma \lambda h (W+X\Omega) - \frac{\beta I}{\mu_0} . 
\ee
As for any function of $\ups$, the quantities $\Sigma$ and $\Lambda$ are conserved along any given fluid line, in consequence of (\ref{e:ups_conserved}). 

Note that if $\dd f \not = 0$, one may choose $\ups = f$, leading to the 
following values [cf. Eqs.~(\ref{e:dPsi_C_df}), (\ref{e:dPhi_D_df}), (\ref{e:E_express}) and
(\ref{e:L_express})]:
\be
  \ups = f \ \Longrightarrow \ \left\{ \begin{array}{l}
  \alpha = D \\
  \beta = C \\
  \gamma = 1 
  \end{array} \right. 
  \ \Longrightarrow \  \left\{ \begin{array}{l}
  \Sigma = E \\
  \Lambda = L . 
  \end{array} \right. 
\ee
Hence, for this choice of $\ups$, $\Sigma$ and $\Lambda$ are simply the Bernoulli-like quantities $E$ and $L$ introduced by BO \cite{BekenO78} and discussed in Sec.~\ref{s:conserv_E_L}. 

If $\dd\Psi\not = 0$, the choice $\ups = \Psi$ is allowed, leading to 
\be \label{e:ups_psi}
  \ups = \Psi \ \Longrightarrow \ \left\{ \begin{array}{l}
  \alpha = -\omega \\
  \beta = 1 \\
  \gamma = C^{-1} 
  \end{array} \right. 
  \ \Longrightarrow \  \left\{ \begin{array}{l}
  \Sigma = E/C \\
  \Lambda = L/C , 
  \end{array} \right. 
\ee
where $\omega = - D/C$ [cf. Eq.~(\ref{e:omega_Omega_I})]. 

\begin{widetext}

\subsection{The master transfield equation}

Having shown that the first two equations of the MHD-Euler system
(\ref{e:syst_MHD_Euler}) leads to the conserved
quantities $\Sigma$ and $\Lambda$, let us focus on the third equation, namely
Eq.~(\ref{e:syst_MHD_Euler-3}). Taking account of (\ref{e:dall_dups}), we can rewrite it as 
\be \label{e:MHD_Euler_ups}
 \lambda \, \dd(h\uu{u}\cdot\vxi)
  +\lambda \Omega \, \dd(h\uu{u}\cdot\vchi) 
  + \frac{1}{n} \left[ \alpha \, \wxis\cdot\vv{j} + \beta \, \wchis\cdot\vv{j}
  - \gamma q - \frac{\gamma \lambda h}{\sigma} \left( \Ci_{\vxi}
   + \Omega \Ci_{\vchi} \right) \right] \, \dd \ups 
   - \frac{I}{\mu_0\sigma n} \, \dd I   + T \, \dd S = 0 . 
\ee
Differentiating expressions~(\ref{e:Sigma_def}) and (\ref{e:Lambda_def})
yields (the prime stands for the first derivative of a function of $\ups$)
\bea
  & & \dd \Sigma = \Sigma' \, \dd\ups = - \gamma \dd(h\uu{u}\cdot\vxi)
  - \left( \gamma' h \uu{u}\cdot\vxi - \frac{\alpha' I}{\mu_0}  \right)
  \dd\ups + \frac{\alpha}{\mu_0} \, \dd I \nonumber \\
  & & \dd \Lambda = \Lambda' \, \dd \ups = \gamma \dd(h\uu{u}\cdot\vchi)
  + \left( \gamma' h \uu{u}\cdot\vchi - \frac{\beta' I}{\mu_0}  \right)\dd\ups
  + \frac{\beta}{\mu_0} \, \dd I , \nonumber
\eea
from which we get
\be \label{e:lduhx}
  \lambda \dd (h\uu{u}\cdot\vxi) + \lambda \Omega \, \dd(h\uu{u}\cdot\vchi) 
  = \frac{\lambda}{\gamma} \left\{
  \left[ \Omega \Lambda' -  \Sigma' + \frac{I}{\mu_0}(\alpha' + \Omega\beta') + \gamma' \lambda h (V - 2W \Omega - X \Omega^2)
 \right] \, \dd \ups
  + \frac{\alpha + \Omega \beta}{\mu_0} \, \dd I \right\}. 
\ee

To treat the entropy term $T\, \dd S$ in Eq.~(\ref{e:MHD_Euler_ups}), let us assume
that $S$ is a function of $\ups$: 
\be \label{e:S_ups}
  S = S(\ups) . 
\ee
Actually (\ref{e:S_ups}) is mandatory if there is a non-vanishing meridional flow. 
Indeed in this case $\dd f \not = 0$ and since $S$ is conserved along the fluid lines [property (\ref{e:S_conserved})], 
Lemma~\ref{l:conserved_f} of Sec.~\ref{s:sec_lemma2} is applicable and
gives $S=S(f)$, i.e., via (\ref{e:all_funct_ups}), $S=S(\ups)$. 
If $\dd f = 0$ (pure rotational motion), then we may consider that 
(\ref{e:S_ups}) is a supplementary hypothesis in our framework, set to integrate the
MHD-Euler equation. Note that a homentropic  fluid ($S=\mathrm{const.}$ throughout the fluid) 
satisfies (\ref{e:S_ups}).  

Substituting Eqs.~(\ref{e:lduhx}) into Eq.~(\ref{e:MHD_Euler_ups}), 
we notice that terms in $\dd I$ cancel each other
thanks to the relation (\ref{e:perfect_alpha_beta}). 
Using (\ref{e:S_ups}) to set $\dd S = S' \, \dd \ups$, 
we are then left with 
\bea
  & & \Bigg\{ \frac{\lambda}{\gamma} 
  \left[ \Omega \Lambda' -  \Sigma' + \frac{I}{\mu_0}(\alpha' + \Omega\beta') + \gamma' \lambda h (V - 2W \Omega - X \Omega^2) \right] \nonumber \\
  & & \qquad + \frac{1}{n} \left[ \alpha \, \wxis\cdot\vv{j} + \beta \, \wchis\cdot\vv{j}
  - \gamma q - \frac{\gamma \lambda h}{\sigma} \left( \Ci_{\vxi}
   + \Omega \Ci_{\vchi} \right) \right]
  + T S' \Bigg\} \, \dd\ups = 0 . \label{e:MHD_Euler_ups_dups}
\eea
Since by hypothesis $\dd\ups \not =0$, 
this equation is equivalent to the vanishing of the term in braces.
Let us express all the pieces in term of $\ups$.  
Writing $\dd \Phi = \alpha \, \dd\ups$  and $\dd\Psi = \beta\,  \dd\ups$
[Eq.~(\ref{e:dall_dups})] in Eqs.~(\ref{e:xis_j}) and (\ref{e:chis_j}), we get 
\bea
  \alpha \, \wxis\cdot\vv{j} + \beta \, \wchis\cdot\vv{j}
 & = & - \frac{1}{\mu_0\sigma} \Big\{ (V\beta^2 + 2 W \alpha\beta - X \alpha^2)
  \Delta^* \ups +  (\beta^2 \dd V + 2\alpha\beta\dd W - \alpha^2 \dd X)\cdot 
  \vec{\wnab}\ups \nonumber \\
  & & + \left[V\beta\beta' + W(\alpha'\beta + \alpha\beta') - X\alpha\alpha'\right]
  \dd \ups\cdot\vec{\wnab}\ups
  - \frac{I}{\sigma} \left[ (W\beta-X\alpha) \Ci_{\vxi}
  + (W\alpha+V\beta) \Ci_{\vchi} \right] \Big\} , \label{e:ajbj}
\eea
where $\Delta^*$ is the operator that generalizes (\ref{e:def_Delta_s_Newt}) to the relativistic
case: 
\be \label{e:def_Delta_s}
  \Delta^* \ups := \sigma \, \nabla_\mu \left( \frac{1}{\sigma} \nabla^\mu \ups \right) . 
\ee
Besides, setting $\dd f = \gamma \,\dd\ups$ [Eq.~(\ref{e:dall_dups})] in 
Eq.~(\ref{e:q_Delta_f}), we have
\be
  q = - \frac{1}{\sigma} \left[ \frac{h\gamma}{n} \Delta^*\ups
  + \gamma \, \dd\left(\frac{h}{n}\right) \cdot\vec{\wnab}\ups
  + \frac{h}{n}\gamma' \, \dd\ups \cdot\vec{\wnab}\ups \right] . \label{e:q_ups}
\ee
Let us substitute Eqs.~(\ref{e:ajbj}) and (\ref{e:q_ups}) into the term in
braces in Eq.~(\ref{e:MHD_Euler_ups_dups}) and express its vanishing. We get,
after multiplication by $\sigma n^2  /h$,
\bea
  & & A \, \Delta^*\ups + \frac{n}{h} \left[ \gamma^2 \dd\left(\frac{h}{n}\right)
  - \frac{1}{\mu_0} \left( \beta^2\dd V + 2\alpha\beta\dd W - \alpha^2 \dd X \right) \right]
   \cdot\vec{\wnab}\ups \nonumber \\
  & &  \qquad + \left\{ \gamma\gamma' - \frac{n}{\mu_0 h} 
  \left[ V \beta\beta' + W (\alpha'\beta + \alpha\beta') - X \alpha\alpha' \right] 
  \right\} \dd \ups \cdot\vec{\wnab}\ups \nonumber \\
  & &  \qquad + 
  \frac{\sigma n^2}{h} \left\{\frac{\lambda}{\gamma} 
  \left[ \Omega \Lambda' -  \Sigma' + \frac{I}{\mu_0}(\alpha' + \Omega\beta') + \gamma' \lambda h (V - 2W \Omega - X \Omega^2) \right]  + T S' \right\} \nonumber \\
  & &  \qquad - \gamma \lambda n  \left( \Ci_{\vxi}
   + \Omega \Ci_{\vchi} \right) 
  + \frac{I n }{\mu_0 \sigma h} \left[ (W\beta-X\alpha) \Ci_{\vxi}
  + (W\alpha+V\beta) \Ci_{\vchi} \right] = 0 , \label{e:master_transfield}
\eea
where 
\be \label{e:def_A}
  A := \gamma^2 - \frac{n}{\mu_0 h} (V\beta^2 + 2 W \alpha\beta - X\alpha^2) . 
\ee
We shall call Eq.~(\ref{e:master_transfield}) the \emph{master transfield equation}. The qualifier \emph{transfield} stems from the fact that it corresponds to the component of the MHD-Euler equation along $\dd \ups$, which is transverse to the magnetic field: $\vv{b}\cdot\dd \ups = 0$, as it is easily verified on the expression 
(\ref{e:b_perfect_cond}) of $\vv{b}$, taking into account 
(\ref{e:all_funct_ups}) and (\ref{e:ups_conserved}). 
Note that the term in $\lambda/\gamma$ in the third line of Eq.~(\ref{e:master_transfield}) is regular, even when $\gamma = 0$, as we shall show
in Sec.~\ref{s:pure_rot}. Besides, 
note that for circular spacetimes (e.g. Kerr spacetime), the last line of
Eq.~(\ref{e:master_transfield}) vanishes identically [cf. 
Eq.~(\ref{e:circ_spacetime})]. 
The master transfield equation was first written in the Newtonian case by Soloviev (1967) \cite{Solov67}, as we shall discuss in Sec.~\ref{s:trans_newt}.

Given the metric [hence the covariant derivative operator $\wnab$,  the coefficients $V$, $W$, $V$ and $\sigma$, 
and the twist scalars $\Ci_{\vxi}$ and $\Ci_{\vchi}$  and the six functions $\alpha(\ups)$, $\beta(\ups)$, $\gamma(\ups)$, $\Sigma(\ups)$, $\Lambda(\ups)$, and $S(\ups)$, the master transfield equation~(\ref{e:master_transfield}) constitutes a (non-linear) second-order PDE for $\ups$. 
Indeed, all the remaining quantities ($I$, $\lambda$, $\Omega$, $n$, $h$, $T$) 
that appear in Eq.~(\ref{e:master_transfield}), although not  functions 
of $\ups$, can be computed once $\ups$ is known, as we will show. 

First of all, by combining Eqs.~(\ref{e:Sigma_def}) and (\ref{e:Lambda_def}), we get
\bea
  & & V \Lambda - W \Sigma =  \gamma \sigma \lambda h \Omega - \frac{I}{\mu_0} (V\beta+W\alpha) \label{e:VL_m_WE} \\
  & & X\Sigma + W\Lambda = \gamma \sigma \lambda h + \frac{I}{\mu_0} (X\alpha - W\beta) . \label{e:XE_p_WL}
\eea
Combining these two equations and using (\ref{e:perfect_alpha_beta}) to express
$\alpha+\Omega \beta$ in terms of $I$, we get 
\be \label{e:I_M_A}
  I = \frac{n}{h A} \left[ (X\alpha-W\beta)\Sigma + (W\alpha + V\beta)\Lambda \right] .
\ee  
Extracting $\lambda h$ from Eq.~(\ref{e:XE_p_WL}) and substituting 
Eq.~(\ref{e:I_M_A}) for $I$ leads to 
\be \label{e:lamb_h_M_A}
  \lambda h = \frac{1}{\sigma \gamma A} \left[ \gamma^2 (X \Sigma + W \Lambda)
  - \frac{\sigma n \beta}{\mu_0 h} (\alpha\Lambda + \beta\Sigma) \right] . 
\ee
Extracting $\Omega$ from Eq.~(\ref{e:VL_m_WE}) and substituting the above values of 
$I$ and $\lambda h$, we obtain the expression of the fluid angular velocities in terms of conserved quantities and $h/n$: 
\be \label{e:Omega_M_A}
  \Omega = \frac{\mu_0 \frac{h}{n} \gamma^2 (V\Lambda-W\Sigma) 
  + \sigma \alpha (\beta\Sigma + \alpha \Lambda)}{ \mu_0 \frac{h}{n} \gamma^2 
  (X\Sigma+W\Lambda) - \sigma \beta (\beta\Sigma+\alpha \Lambda)} . 
\ee

Besides, from the relation (\ref{e:w_df}), we have 
\be \label{e:w2_df2}
  \uu{w}\cdot\vv{w} = \frac{1}{\sigma n^2} \, \dd f \cdot \vec{\wnab} f
  = \frac{\gamma^2}{\sigma n^2} \, \dd \ups \cdot \vec{\wnab} \ups , 
\ee
so that the 4-velocity normalization relation (\ref{e:lambda}) can be written as 
\be
  1 + \frac{\gamma^2}{\sigma n^2} \, \dd \ups \cdot \vec{\wnab} \ups = \lambda^2 (V - 2W \Omega - X \Omega^2) . 
\ee
Substituting Eq.~(\ref{e:Omega_M_A}) for $\Omega$ and 
writing $\lambda = (\lambda h)/h$ with $\lambda h$ given by (\ref{e:lamb_h_M_A}), we get, after some rearrangements, 
\be \label{e:master_wind_prov}
  h^2 \left( \sigma + \frac{\gamma^2}{n^2} \, 
  \dd\ups \cdot\vec{\wnab}\ups \right) - \frac{\gamma^2}{A^2} ( X\Sigma^2 + 2W \Sigma\Lambda
  - V\Lambda^2) + \frac{\sigma n}{\mu_0 h} \frac{A+\gamma^2}{A^2\gamma^2}
  (\beta\Sigma + \alpha \Lambda)^2 = 0 . 
\ee
By means of the identity
\[
  \sigma (\beta\Sigma + \alpha \Lambda)^2 = \left[ (X\alpha - W\beta) \Sigma
  + (V\beta + W\alpha) \Lambda \right]^2 
  + (V\beta^2 + 2 W \alpha\beta - X \alpha^2) ( X\Sigma^2 + 2W \Sigma\Lambda
  - V\Lambda^2), 
\]
which follows solely from $\sigma = XV + W^2$, Eq.~ (\ref{e:master_wind_prov}) 
can be recast in the alternative form\footnote{A combination 
$-\Sigma + \Omega \Lambda$ of Eqs.~(\ref{e:Sigma_def}) and (\ref{e:Lambda_def})
may be used to derive Eqs.~(\ref{e:master_wind_prov}) and (\ref{e:master_wind}) 
from $\uu{u}\cdot\vv{u}=-1$. }
\be \label{e:master_wind}
  h^2 \left( \sigma + \frac{\gamma^2}{n^2} \, 
  \dd\ups \cdot\vec{\wnab}\ups \right)
  - \frac{1}{\gamma^2} \left( X\Sigma^2 + 2W \Sigma\Lambda
  - V\Lambda^2 \right)
  + \frac{n}{\mu_0 h} \frac{A+\gamma^2}{A^2\gamma^2}
  \left[ (X\alpha - W\beta) \Sigma
  + (V\beta + W\alpha) \Lambda \right]^2 = 0 . 
\ee
This equation is called the \emph{poloidal wind equation}. 
Given the metric factors $V$, $W$, $X$ and $\sigma$, 
the functions $\alpha(\ups)$, $\beta(\ups)$, $\gamma(\ups)$, $\Sigma(\ups)$, $\Lambda(\ups)$ and $S(\ups)$, expression (\ref{e:def_A}) for $A$, as well as the EOS
$h = h(s,n)$ with $s=S(\ups)n$ [cf. Eq.~(\ref{e:def_S})],
the poloidal wind equation can be solved to compute
$n$ once $\ups$ is known. 
Then, from $n$ we get $h$ via the EOS
and $A$ via Eq.~(\ref{e:def_A}). 
Once $n$, $h$ and $A$ are known, we can compute $I$ via Eq.~(\ref{e:I_M_A})
and $\Omega$ via Eq.~(\ref{e:Omega_M_A}). The 
meridional velocity 
$\vv{w}$ is obtained via Eq.~(\ref{e:w_df}) with $\dd f = \gamma \, \dd \ups$
and the velocity coefficient $\lambda$ via Eq.~(\ref{e:lambda}). 

An equivalent point of view is to consider that the fundamental equations to be solved
are Eqs.~(\ref{e:master_transfield}) and (\ref{e:master_wind}) which constitute a 
coupled PDE system for the
two unknowns $(\ups,n)$. Indeed, given the metric, the EOS and the six functions 
$\alpha(\ups)$, $\beta(\ups)$, $\gamma(\ups)$, $\Sigma(\ups)$, $\Lambda(\ups)$ and $S(\ups)$, solving this system provides a solution of the 
MHD-Euler equation and Maxwell equations, the electromagnetic field tensor $\w{F}$ and
electric 4-current $\vv{j}$
being deduced from $\ups$ via Eqs.~(\ref{e:F_gal}), (\ref{e:4current}), (\ref{e:dall_dups}) and (\ref{e:I_M_A}).


\section{Subcases of the master transfield equation} \label{s:subcases}

The master transfield equation (\ref{e:master_transfield}), coupled with the 
poloidal wind equation (\ref{e:master_wind}), describes the most general MHD equilibria in generic (noncircular) stationary and axisymmetric spacetimes. 
We shall now specialize it to various cases and make the link with results obtained previously in the literature. 

\subsection{Newtonian limit} \label{s:trans_newt}

At the Newtonian limit, as given by Eqs.~(\ref{e:Newt}), (\ref{e:circ_spacetime}),
(\ref{e:h_H}) and (\ref{e:lambda_Newt}), the expressions
(\ref{e:Sigma_def}) and (\ref{e:Lambda_def}) of the streamline-conserved quantities 
$\Sigma(\ups)$ and $\Lambda(\ups)$ reduce to 
\bea
  \Sigma & = & \gamma m_{\rm b} \left( 1 + H + \Phi_{\rm grav} + \frac{v^2}{2} \right)
    + \frac{\alpha I}{\mu_0} \label{e:Sigma_Newt}\\
  \Lambda & = & \gamma m_{\rm b} r^2\sin^2\theta \Omega - \frac{\beta I}{\mu_0} ,
  \label{e:Lambda_Newt}
\eea
whereas the master transfield 
equation (\ref{e:master_transfield}) reduces to
\bea
  & & A \Delta^*\ups - \frac{\gamma^2}{n} \, \dd n \cdot \vec{\wnab}\ups
  + \left( \gamma \gamma' - \frac{n}{\mu_0 m_{\rm b}} \beta\beta' \right)
  \dd\ups \cdot \vec{\wnab}\ups \nonumber \\
  & & \qquad + r^2\sin^2\theta \, \frac{n^2}{m_{\rm b}} \left\{\frac{1}{\gamma} 
  \left[ \Omega \Lambda' -  \Sigma' + \frac{I}{\mu_0}(\alpha' + \Omega\beta') + \gamma' m_{\rm b} \right]  + T S' \right\} = 0 , \label{e:transfield_Newt}
\eea
with the expression (\ref{e:def_Delta_s_Newt}) for the operator $ \Delta^*$ and\footnote{Note that in taking the Newtonian limit of (\ref{e:def_A}), the term
$X \alpha^2 = r^2\sin^2\theta \alpha^2$ is relativistic and therefore must be
disregarded.}
\be
  A =\gamma^2 - \frac{n\beta^2}{\mu_0 m_{\rm b}} . 
\ee
Performing the proper changes of notation\footnote{The link between Soloviev notations \cite{Solov67} and ours is $r\leftrightarrow r\sin\theta$, $s\leftrightarrow m_{\rm b} A/n$, $\Psi'_0 \leftrightarrow  m_{\rm b} \gamma$, $\Psi'\leftrightarrow \beta / \sqrt{\mu_0}$, 
$I_0 \leftrightarrow \Omega r^2\sin^2\theta$, $I \leftrightarrow I / \sqrt{\mu_0}$,
$S \leftrightarrow S/m_{\rm b}$, 
$A\leftrightarrow \Lambda$, $B\leftrightarrow \alpha/\sqrt{\mu_0}$,  
and $U\leftrightarrow (m_{\rm b}\gamma)^{-1} (\Sigma + \alpha \Lambda/\beta) -1$.}, 
one can check that 
Eq.~(\ref{e:transfield_Newt}) coincides with the
first equation in the system~(II) of Soloviev \cite{Solov67}. 
After Soloviev work, the Newtonian transfield equation has been re-obtained by many authors
for the special case in which $\ups = \Psi$ [cf. (\ref{e:ups_psi})]
(e.g. \cite{Okamo75}, \cite{Tsing82}, \cite{Sakur85}, \cite{LovelMMS86}, \cite{HeyvaN89}; 
cf. \cite{Heyva96} for an extended discussion and \cite{DuezM10} for a recent study).  
The equation is then known as the \emph{generalized Grad-Shafranov equation}
(see Sec.~\ref{s:GS} below).  

The Newtonian limits of expressions (\ref{e:I_M_A}) and (\ref{e:Omega_M_A}) for $I$ 
and $\Omega$ are 
\bea
  I & = & \frac{n}{A} \left( \alpha \gamma r^2\sin^2\theta  + \frac{\beta\Lambda}{m_{\rm b}}
  \right) , \label{e:I_M_A_Newt} \\
  \Omega & = & \frac{1}{m_{\rm b} A} \left( \frac{\gamma\Lambda}{r^2\sin^2\theta}
  + \frac{n}{\mu_0} \alpha\beta \right) . \label{e:Omega_M_A_Newt}
\eea
To get the latter expression, we have used the Newtonian limit 
$\beta\Sigma + \alpha\Lambda = m_{\rm b} \beta\gamma$, resulting from 
(\ref{e:Sigma_Newt})-(\ref{e:Lambda_Newt}). 
Equations~(\ref{e:I_M_A_Newt}) and (\ref{e:Omega_M_A_Newt}) 
coincides with respectively the second and first equations in Eq.~(1.18) of Soloviev 
article \cite{Solov67}. 

Finally to get the Newtonian limit of the
poloidal wind equation (\ref{e:master_wind}), 
we rewrite the term $h^2 \sigma - X \Sigma^2 / \gamma^2$ by means of  
the Newtonian expressions (\ref{e:Newt}) of $\sigma$
and $X$, along with (\ref{e:h_H}):
\[
  \sigma h^2 - \frac{X\Sigma^2}{\gamma^2} 
 \simeq r^2\sin^2\theta \left[ m_{\rm b}^2 (1+2H)
  - (1-2\Phi_{\rm grav}) \frac{\Sigma^2}{\gamma^2} \right]
  \simeq 2 m_{\rm b}^2 r^2\sin^2\theta \left[ H + \Phi_{\rm grav}
    + 1-\frac{\Sigma}{m_{\rm b}\gamma}
\right] , 
\]
where the last equality results from 
$1-(\Sigma/ m_{\rm b}\gamma)^2 = (1+\Sigma/ m_{\rm b}\gamma)(1-\Sigma/ m_{\rm b}\gamma) 
\simeq 2 (1-\Sigma/ m_{\rm b}\gamma)$.
Accordingly, 
the Newtonian limit of the poloidal wind equation~(\ref{e:master_wind}) is
\be \label{e:master_wind_Newt}
  \frac{\gamma^2}{n^2} \dd\ups\cdot\vec{\wnab}\ups  + 2 r^2\sin^2\theta 
  \left( H + \Phi_{\rm grav} + 1 - \frac{\Sigma}{m_{\rm b}\gamma} \right)
  +\left( \frac{\Lambda}{m_{\rm b}\gamma} \right) ^2 
+ \frac{n}{\mu_0 m_{\rm b}} \frac{A+\gamma^2}{A^2} 
  \left( \frac{\beta\Lambda}{m_{\rm b}\gamma} + \alpha r^2\sin^2\theta  \right) ^2 = 0 . 
\ee
This equation is not exhibited in Soloviev work \cite{Solov67}. It can however be recovered
by combining Soloviev Eqs.~(1.5) and (1.25) and expressing $v^2$ as
$\Omega^2 r^2\sin^2\theta + \uu{w}\cdot\vv{w}$ with $\Omega$ substituted by 
(\ref{e:Omega_M_A_Newt}) and $\uu{w}\cdot\vv{w}$ by (\ref{e:w2_df2}). 
In the special case where $\ups=\Psi$, one can check that Eq.~(\ref{e:master_wind_Newt})
coincides with Eq.~(14) of Heyvaerts and Norman \cite{HeyvaN89}
(called the \emph{Bernoulli equation} by these authors). 

\end{widetext}

\subsection{Pure rotational flow} \label{s:rot_flow}

The case of a pure rotational flow corresponds to 
\be
  \vv{w} = 0 \iff \dd f = 0 \iff \gamma = 0 . 
\ee
Then Eq.~(\ref{e:perfect_alpha_beta}) yields
\be \label{e:al_Omega_bet}
  \alpha = - \Omega \beta , 
\ee
whereas Eqs.~(\ref{e:Sigma_def}) and (\ref{e:Lambda_def}) reduce to 
\be \label{e:Sigma_Lambda_rot}
  \Sigma = \frac{\alpha I}{\mu_0} \qquad\mbox{and}\qquad 
  \Lambda = - \frac{\beta I}{\mu_0} . 
\ee
If $\alpha\not =0$ (i.e. $\dd\Phi\not=0$) or $\beta\not=0$ (i.e. $\dd\Psi\not=0$), 
Eqs.~(\ref{e:al_Omega_bet}) and (\ref{e:Sigma_Lambda_rot}) imply that 
$\Omega$ and $I$ are functions of $\ups$ (for $\alpha$, $\beta$, $\Sigma$ and $\Lambda$
are all functions of $\ups$):
\be \label{e:Omega_I_rot}
  \Omega = \Omega(\ups) \qquad\mbox{and}\qquad I = I(\ups) . 
\ee
Taking into account Eq.~(\ref{e:al_Omega_bet}) and $\gamma=0$, 
the expression (\ref{e:def_A}) for $A$ becomes
\be \label{e:A_rot}
  A = - \frac{\beta^2 n}{\mu_0 h} (V - 2W\Omega - X \Omega^2) . 
\ee

To express the master transfield equation (\ref{e:master_transfield}) in the case
$\gamma=0$, we shall first evaluate the term which is divided by $\gamma$ in 
Eq.~(\ref{e:master_transfield}), namely 
\be \label{e:def_mA}
  \mathcal{A} := 
\Omega \Lambda' -  \Sigma' + \frac{I}{\mu_0}(\alpha' + \Omega\beta') + \gamma' \lambda h (V - 2W \Omega - X \Omega^2) . 
\ee
To this aim, we shall first suppose $\gamma\not = 0$ and, in a second stage,
take the limit $\gamma\rightarrow 0$. Since $I=I(\ups)$ when $\gamma=0$ [Eq.~(\ref{e:Omega_I_rot})], we 
may write
\be \label{e:I_I_0_a}
  I = I_0(\ups) + \gamma a , 
\ee
where $I_0(\ups)$ is a function of $\ups$ and $a$ describes the behavior of $I$ as
$\gamma\rightarrow 0$. For instance, if $\beta\not=0$ (i.e. $\dd\Psi\not=0$), explicit values of $I_0$ and $a$ are deduced from Eq.~(\ref{e:Lambda_def}):
\[
  I_0(\ups) = - \mu_0 \frac{\Lambda(\ups)}{\beta(\ups)}
  \quad\mbox{and}\quad
  a = \mu_0 \frac{\lambda h}{\beta}(W+X\Omega) . 
\]
Substituting expression (\ref{e:I_I_0_a}) for $I$ into 
Eqs.~(\ref{e:Sigma_def}) and (\ref{e:Lambda_def}), we get
\be \label{e:SL_GH}
  \Sigma = \frac{\alpha}{\mu_0} I_0 + \gamma G 
  \quad\mbox{and}\quad
  \Lambda = - \frac{\beta}{\mu_0} I_0 + \gamma H , 
\ee
with
\be \label{e:G_H_def}
  G := \lambda h (V-W\Omega) + \frac{\alpha a}{\mu_0}
   \quad\mbox{and}\quad
  H := \lambda h (W + X\Omega) - \frac{\beta a}{\mu_0} . 
\ee
From Eq.~(\ref{e:SL_GH}) and the fact that $\Sigma$, $\Lambda$, $\alpha$, $\beta$, $\gamma$ and $I_0$ are all functions of $\ups$, it is clear that $G=G(\ups)$ and $H=H(\ups)$.
Then, using successively Eqs.~(\ref{e:SL_GH}), (\ref{e:G_H_def}) and (\ref{e:perfect_alpha_beta}),
we may write expression (\ref{e:def_mA}) as
\be \label{e:mA_prov}
  \mathcal{A} = \gamma \left\{ \Omega H' - G' + \frac{1}{\mu_0}
  \left[ a(\alpha'+ \Omega \beta') - \frac{I}{\sigma n\lambda} (I'_0 + a \gamma') \right]
  \right\} . 
\ee
Taking the limit $\gamma\rightarrow 0$, we have $\Omega = \Omega(\ups)$
[Eq.~(\ref{e:Omega_I_rot})] and $\alpha = - \Omega \beta$ [Eq.~(\ref{e:al_Omega_bet})], 
so that $\alpha' + \Omega \beta' = - \Omega' \beta$ and 
$\Omega H' - G' = - K' - \Omega' H$ with $K(\ups) := G - \Omega H$. 
According to (\ref{e:G_H_def})
and (\ref{e:al_Omega_bet}), we have $K =\lambda h (V-2W \Omega - X\Omega^2)$. 
Now for $\gamma=0$, expression (\ref{e:lambda}) for $\lambda$ reduces to
\be \label{e:lambda_rot}
\lambda = (V-2W \Omega - X \Omega^2)^{-1/2} ,  
\ee
so that 
\be \label{e:K_h}
  K(\ups) = \frac{h}{\lambda} = h \sqrt{V-2W \Omega - X\Omega^2} . 
\ee
Since
\begin{eqnarray}
  \Omega H' - G' + \frac{a}{\mu_0}(\alpha'+ \Omega \beta') 
  &=& -K' - \Omega' H   - \frac{a}{\mu_0} \Omega' \beta \nonumber \\  
  &=& - K' - \lambda h (W+X\Omega) \Omega',  \nonumber
\end{eqnarray}
the limit $\gamma\rightarrow 0$ of Eq.~(\ref{e:mA_prov}) is
\be \label{e:lim_Asg}
  \lim_{\gamma \rightarrow 0} \frac{\mathcal{A}}{\gamma} 
  = - K' 
  - \lambda h (W+X\Omega) \Omega' - \frac{I I'}{\mu_0\sigma n\lambda} .
\ee

\begin{widetext}

Thanks to the relations (\ref{e:al_Omega_bet}), (\ref{e:A_rot}), (\ref{e:lambda_rot}),
(\ref{e:lim_Asg}) and (\ref{e:def_h}), the master transfield equation (\ref{e:master_transfield}) becomes
\bea
  & & \frac{\beta^2}{\lambda^2} \, \Delta^*\ups
  + \beta \, \dd\left( \frac{\beta}{\lambda^2} \right) \cdot\vec{\wnab}\ups
  + \beta^2 (W+X\Omega) \Omega' \, \dd\ups\cdot\vec{\wnab}\ups 
  + I \left\{ I' - \frac{\beta}{\sigma} \left[ (W+X\Omega) \Ci_{\vxi}
  + (V-W\Omega) \Ci_{\vchi} \right] \right\} \nonumber \\
  & & \qquad + \mu_0\sigma\left\{ (\vep + p) \left[ 
  \frac{K'}{K}
  + \lambda^2 (W+X\Omega)\Omega' \right] - n T S' \right\}= 0 . \label{e:transfield_rot}
\eea

Let us now take the pure rotational limit of the wind equation, in
the form (\ref{e:master_wind_prov}).
From Eq.~(\ref{e:SL_GH}), $\beta\Sigma + \alpha \Lambda = \gamma (\beta G + \alpha H)$,
so that, in view of (\ref{e:al_Omega_bet}),
\[
  \lim_{\gamma \rightarrow 0} \frac{1}{\gamma} (\beta\Sigma + \alpha \Lambda) = 
  \beta (G-\Omega H) = \beta K . 
\]
This property, along with (\ref{e:A_rot}), implies that 
for $\gamma = 0$ the wind equation (\ref{e:master_wind_prov}) reduces to 
\[
  h^2 - \frac{K^2}{V - 2 W \Omega - X\Omega^2} = 0 , 
\]
which is nothing but the square of Eq.~(\ref{e:K_h}). Consequently, the wind equation is
trivially satisfied in this case. 

In conclusion, for a pure rotational flow, one should prescribe five functions of
the master potential: $\beta(\ups)$, $\Omega(\ups)$, $I(\ups)$, $K(\ups)$ 
and $S(\ups)$ and solve
for the transfield equation (\ref{e:transfield_rot}) for $\ups$. In that equation, the matter quantities $\vep$, $p$, $n$ and $T$ are given by the EOS from the knowledge of
$S$ and $h$, the latter being deduced from $\Omega(\ups)$ and  $K(\ups)$ 
via Eq.~(\ref{e:K_h}). We shall discuss further the pure rotational flow below
(Sec. \ref{s:pure_rot} and \ref{s:tor_rot}). 

\subsection{Expression in terms of $\Psi$: generalized Grad-Shafranov equation}
\label{s:GS}

Let us assume that $\dd\Psi\not =0$. We may then choose $\ups = \Psi$ as the primary variable [cf. (\ref{e:ups_psi})]. This is actually the choice performed by most 
(all ?) of previous relativistic studies, disregarding the case
$\dd\Psi=0$ (toroidal magnetic field or hydrodynamical 
limit, to be discussed in Sec.~\ref{s:toroidal} and \ref{s:hydro_lim}). 
Let us first consider the pure rotational flow, in order to make the link with the original Grad-Shafranov equation. 

\subsubsection{Pure rotational flow} \label{s:pure_rot}

For a pure rotational flow with $\ups=\Psi$, Eq.~(\ref{e:Omega_I_rot}) become
\be \label{e:Omega_I_Psi}
  \Omega = \Omega(\Psi) \qquad\mbox{and}\qquad I = I(\Psi) . 
\ee
In the Newtonian regime, the property $\Omega = \Omega(\Psi)$ is known as \emph{Ferraro's law of isorotation} \cite{Ferra37}, while the result $I=I(\Psi)$ has been obtained by Chandrasekhar (1956) \cite{Chandr56}. 
The transfield equation (\ref{e:transfield_rot}) becomes [cf. (\ref{e:ups_psi})
and (\ref{e:lambda_rot})]
\bea
  & & (V-2W\Omega-X\Omega^2) \Delta^* \Psi +   
  \dd (V-2W\Omega-X\Omega^2)\cdot \vec{\wnab} \Psi 
  + (W+X\Omega)  \Omega' \, \dd \Psi \cdot \vec{\wnab} \Psi
+ I \left[ I' - \frac{W+X\Omega}{\sigma} \Ci_{\vxi}
- \frac{V-W\Omega}{\sigma} \Ci_{\vchi} \right] \nonumber \\
  & & \ \qquad + \mu_0 \sigma \left\{ (\vep+p) \left[ \frac{K'}{K}
 + \frac{(X\Omega+W) \Omega'}{V-2W\Omega-X\Omega^2} \right]
  -  n T S' \right\} = 0 .   \label{e:GS_Psi}
\eea
This PDE has to be solved for $\Psi$, once the four functions $\Omega(\Psi)$, 
$I(\Psi)$, $K(\Psi)$ 
and $S(\Psi)$ are prescribed. The enthalpy field $h$ is given by Eq.~(\ref{e:K_h})
which remains unchanged. 

Equation~(\ref{e:GS_Psi}) is a relativistic generalization of the so-called \emph{Grad-Shafranov} 
equation \cite{GradR58,Grad60,Shafr58,Shafr66} (see also Chap.~16 of the recent textbook 
\cite{GoedbKP10}). 
At the Newtonian limit [cf. expressions (\ref{e:Newt}) and (\ref{e:def_rho_H})] and in coordinates
$(t,r,\theta,\ph)$ of spherical type, Eq.~(\ref{e:GS_Psi})
reduces to 
\be \label{e:GS_Psi_Newt}
        \mbox{\footnotesize Newt.:}\qquad  \Delta^* \Psi + I I' + \mu_0 r^2 \sin^2\theta \left[ 
        \rho  \left( K'/m_{\rm b} + \Omega \Omega'r^2\sin^2\theta  \right) - n T S' \right] = 0 , 
\ee
whereas Eq.~(\ref{e:K_h}) reduces to [cf. (\ref{e:h_H}) and (\ref{e:lambda_Newt})]
\be \label{e:first_integ_rot_Psi_Newt}
        \mbox{\footnotesize Newt.:}\qquad  H + \Phi_{\rm grav} - \frac{1}{2} \Omega^2 r^2 \sin^2 \theta = \frac{K(\Psi)}{m_{\rm b}} - 1  .
\ee
Let us recall that the Newtonian expression of $\Delta^*$ is given by Eq.~(\ref{e:def_Delta_s_Newt}). 
Modulo the change from spherical to cylindrical coordinates,
Eqs.~(\ref{e:GS_Psi_Newt}) and (\ref{e:first_integ_rot_Psi_Newt}) coincide with 
respectively Eqs.~(3.3) and (3.2) of Maschke and Perrin \cite{MaschP80}.
The limiting case $I=0$ [pure poloidal magnetic field, cf. Eq.~(\ref{e:b_Newt})], 
$\rho = \mathrm{const}$, 
$\Omega=\mathrm{const}$
and $S=\mathrm{const}$ has been treated by Ferraro in 1954 \cite{Ferra54}. 
It has been extended to $I\not=0$ and $\Omega\not=\mathrm{const}$, still maintaining 
$\rho=\mathrm{const}$, 
by Chandrasekhar in 1956 \cite{Chandr56} (using the function $P:= \Psi / (r^2\sin^2\theta)$ instead of $\Psi$). 
Plasma physicists Grad and Rubin \cite{GradR58} and Shafranov \cite{Shafr58} have considered in 1958 the non-rotating limit ($\Omega=0$) of Eq.~(\ref{e:GS_Psi_Newt}) (see Chap.~16 of \cite{GoedbKP10}). 

Coming back to the relativistic case, the special case $I=0$, $\Omega = \mathrm{const}$
and $S = \mathrm{const}$ has been discussed by Bonazzola et al. \cite{BonazGSM93} and
Bocquet et al. \cite{BocquBGN95}. Note that contrary to what is claimed in 
Ref.~\cite{BonazGSM93},  $\Omega$ has not to be a constant: it can be any function of $\Psi$
[Eq.~(\ref{e:Omega_I_Psi})]. 
Most relativistic studies have focused
on the case $\vv{w}\not=0$ (flow with a meridional component) and barely discussed the 
limit $\vv{w}=0$ presented above. In particular, it is claimed in Ref.~\cite{IokaS03} that
if $\vv{w}\rightarrow 0$, the magnetic field $\vv{b}$ cannot have a toroidal component (i.e. $I=0$). We see no support of this since any choice seems to be allowed for the function 
$I(\Psi)$ in the equations presented above.

\subsubsection{Generic flow}

For a generic flow (i.e. with some meridional component), we have, according to 
(\ref{e:ups_psi}), $\alpha = -\omega$,
$\beta = 1$ and 
$\gamma = C^{-1}$, with $C\not=0$ since $\dd \Psi\not=0$. The expression 
(\ref{e:def_A}) for $A$ becomes then 
\be
  A = \frac{1}{C^2} \left( 1 - \frac{V-2W\omega - X\omega^2}{M^2} \right), 
\ee
where $M$ is the \emph{poloidal Alfv\'en Mach-number} : 
\be \label{e:def_M_A}
  M^2 := \frac{\mu_0 h}{C^2 n} . 
\ee
This name is justified by the Newtonian limit [cf. (\ref{e:h_H})]: 
\be \label{e:Mach_Newt}
  \mbox{\footnotesize Newt.:}\quad  M^2 = \frac{\mu_0 m_{\rm b}}{C^2 n} = \left( \frac{|\vv{w}|}{v_{\rm A,p}}
  \right) ^2 ,
  \qquad v_{\rm A,p} := \frac{|\vv{b}_{\rm p}|}{\sqrt{\mu_0 n m_{\rm b}}} . 
\ee
$v_{\rm A,p}$ is the \emph{poloidal Alfv\'en velocity}, $\vv{b}_{\rm p}$ being 
the poloidal magnetic field [cf. Eq.~(\ref{e:b_p_colin_w})]. The expression
of $M$ as the ratio of the norm of $\vv{w}$ to $v_{\rm A,p}$ justifies the name \emph{poloidal Alfv\'en Mach-number}
given to $M$.

Setting $\Sigma = E/C$ and $\Lambda = L/C$ [cf. (\ref{e:ups_psi})],
the transfield equation (\ref{e:master_transfield}) specialized to $\ups=\Psi$ is
\bea
  && \left( 1 - \frac{V-2W\omega - X\omega^2}{M^2} \right) \Delta^* \Psi
  + \left[ \frac{n}{h}  \dd\left(\frac{h}{n}\right) 
  - \frac{1}{M^2} \left( \dd V - 2\omega\dd W - \omega^2 \dd X\right)
  \right] \cdot \vec{\wnab} \Psi  
   \nonumber \\
&& \qquad + \left[ \frac{\omega'}{M^2} (W+X\omega) - \frac{C'}{C} \right] \, \dd\Psi \cdot \vec{\wnab} \Psi 
  + \frac{\mu_0 \sigma n}{M^2} \left\{ \lambda \left[
  \Omega L' - E' + \frac{I}{\mu_0} \left( C'(\Omega -\omega) 
   -C\omega' \right) \right] + T S' \right\} 
  \nonumber \\
&& \qquad- \lambda n C \left( \Ci_{\vxi}
   + \Omega\,  \Ci_{\vchi} \right) + \frac{I}{\sigma M^2}
  \left[(W+X\omega)  \Ci_{\vxi}+ (V-W\omega) \Ci_{\vchi} \right] = 0 . 
  \label{e:gen_GS}
\eea
In this equation, all the primes denote derivatives with respect to $\Psi$. 
Equation~(\ref{e:gen_GS}) is called the 
\emph{generalized Grad-Shafranov equation}, since it can be considered as an 
extension of the Grad-Shafranov equation (\ref{e:GS_Psi}) to the case of a non-vanishing meridional flow. 
The generalized Grad-Shafranov equation has been derived for Minkowski spacetimes by 
Camenzind (1987) \cite{Camen87} and Heyvaerts \& Norman (2003) \cite{HeyvaN03}. It has been extended 
to weak gravitational fields by Lovelace et al. (1986) \cite{LovelMMS86} and to  the
Schwarzschild spacetime by Mobarry and Lovelace (1986) \cite{MobarL86}. 
The case of the Kerr spacetime has been first considered by Nitta et al. (1991)
\cite{NittaTT91} for pressureless matter and Beskin and Pariev (1993) \cite{BeskiP93} for 
non-vanishing pressure (see also \cite{Beski10}). Finally the general case of noncircular stationary axisymmetric spacetimes has been treated by Ioka and Sasaki (2003) \cite{IokaS03,IokaS04}. Note however that they have not written the generalized Grad-Shafranov equation explicitly as Eq.~(\ref{e:gen_GS}), but have kept 
the $\wxis\cdot\vv{j}$, $\wchis\cdot\vv{j}$ and $q$ terms, as in Eq.~(\ref{e:MHD_Euler_ups_dups}).
They have replaced these terms, leading to $\Delta^* \Psi$, only when taking the Newtonian limit. In particular, it is not apparent in their work that the Grad-Shafranov equation
is singular at the Alfv\'en surface, where the term in factor of $\Delta^*\Psi$ vanishes
(see below).  
Besides, as stated in the Introduction, their approach appeals to a (2+1)+1 
foliation of spacetime, whereas ours does not require any extra structure. 

Regarding the poloidal wind equation (\ref{e:master_wind}), it reads for $\Psi=\ups$, 
\be \label{e:wind_psi}
  \frac{1}{C^2 n^2}  \, \dd \Psi \cdot \vec{\wnab} \Psi +  \sigma h^2
  - XE^2 - 2WEL + VL^2 + \frac{2{\tilde M}^2 -1}{({\tilde M}^2-1)^2} \, 
  \frac{\left[ (V-W\omega) L - (W+X\omega)E \right] ^2}{V-2W\omega - X\omega^2 } = 0 ,  
\ee
where\footnote{Note that ${\tilde M}^2$ is not necessarily positive, contrary to 
$M^2$.}
\be \label{e:def_tilde_M}
  {\tilde M}^2 := \frac{M^2}{V-2W\omega - X\omega^2} = \frac{\mu_0}{C^2(V-2W\omega - X\omega^2)} \, \frac{h}{n} . 
\ee
Equations~(\ref{e:gen_GS}) and (\ref{e:wind_psi}) form a system of two equations for 
the two unknowns $(\Psi,n)$, once the five functions $\omega(\Psi)$, $C(\Psi)$, 
$E(\Psi)$, $L(\Psi)$ and $S(\Psi)$ are prescribed, as well as the EOS. In these
equations, $I$ is expressed via Eq.~(\ref{e:I_M_A}):
\be \label{e:I_psi}
  I = \frac{\mu_0}{C} \, \frac{(V-W\omega) L - (W+X\omega) E}{M^2 - V^2 + 2 W\omega + X\omega^2}
\ee
and $\Omega$ via Eq.~(\ref{e:Omega_M_A}) recast as 
\be \label{e:Omega_psi}
  \Omega = \frac{M^2(VL-WE) - \sigma \omega (E-\omega L)}{M^2(XE+WL) - \sigma(E-\omega L)} . 
\ee
As a check, one may verify that Eq.~(\ref{e:I_psi}) coincides
with Eq.~(24) obtained by Beskin \& Pariev \cite{BeskiP93} for 
the Kerr spacetime\footnote{The link between notations of Ref.~\cite{BeskiP93} and
ours is as follows: $\alpha^2 \leftrightarrow \sigma/X$, $\omega \leftrightarrow - W/X$,
$\gamma^2 \leftrightarrow \sigma \lambda^2 /X$,  
$\varpi^2 \leftrightarrow X$, $I\leftrightarrow -I/2$,
$\Psi\leftrightarrow \mu_0 \Psi/2$, $\eta \leftrightarrow C^{-1}$, 
$\Omega^F\leftrightarrow \omega$,  
$E\leftrightarrow E/C$ and $L \leftrightarrow L/C$.} and that Eq.~(\ref{e:Omega_psi}) coincides\footnote{The link between notations of 
Ref.~\cite{Camen86a} and ours is $\alpha \leftrightarrow \omega$ and
$R^2 \leftrightarrow \sigma$.} with Eq.~(110) obtained by Camenzind \cite{Camen86a}
for the Minkowski spacetime. It can also be recovered for general circular spacetimes by combining
Eqs.~(38), (39) and (41) of Ref.~\cite{Camen86b}.

The generalized Grad-Shafranov equation (\ref{e:gen_GS}) is singular for $M^2 = V-2W\omega - X\omega^2$, or equivalently, for 
${\tilde M}^2 = 1$. This condition defines the so-called \emph{Alfv\'en surface}
(see e.g. Refs.~\cite{Beski10,Camen86a,Camen86b,Beski97,TakahNTT90} for an extended discussion). 
The term ${\tilde M}^2 - 1$ also appears at the denominator in the
the poloidal wind equation
(\ref{e:wind_psi}) or in expression (\ref{e:I_psi}) for $I$, but this does not make these equations
singular at the Alfv\'en surface, thanks to the simultaneous vanishing of the 
corresponding numerator \cite{Beski10}. 

\subsection{Toroidal magnetic field ($\mathrm{d}\Psi=0$)} \label{s:toroidal}
 
The complementary case of that treated in the previous subsection is 
\be
  \dd\Psi = 0 \iff \beta = 0 ,
\ee 
where the equivalence follows from the very definition of $\beta$ given in
Eq.~(\ref{e:dall_dups}).
Then, the expression (\ref{e:b_perfect_cond}) for the magnetic field in the fluid frame reduces
to 
\be \label{e:b_quasi_toroidal}
        \vv{b} =  \frac{\lambda I}{\sigma} \left[
   (W+X \Omega) \, \vxi  + \left( V-W \Omega - \frac{\uu{w}\cdot\vv{w}}{\lambda^2}\right)
        \,  \vchi  - \frac{1}{\lambda}(W+X \Omega) \, \vv{w} \right].
\ee
Strictly speaking, this field is not purely toroidal, except at the
Newtonian limit or when $\vv{w}=0$. 
By a slight abuse of language, we shall however
refer to the case $\dd\Psi=0$ as the \emph{toroidal magnetic field case}. 

\subsubsection{Generic case} \label{s:tor_gen}

With $\beta=0$, the perfect conductivity relation (\ref{e:perfect_alpha_beta}) reduces to 
\be \label{e:alpha_tor}
  \alpha =  \frac{\gamma I}{\sigma n\lambda} .
\ee
Consequently, the expression (\ref{e:def_A}) of $A$ becomes
\be
  A = \gamma^2 \left( 1 + \frac{X I^2}{\mu_0 \sigma^2 \lambda^2 n h } \right) . 
\ee

The master transfield equation (\ref{e:master_transfield}) reduces to 
\bea
  & & A \, \Delta^*\ups + \gamma^2 \frac{n}{h} \left[ \dd\left(\frac{h}{n}\right)
   + \frac{I^2}{\mu_0 \sigma^2 \lambda^2 n h}  \dd X  \right]
   \cdot\vec{\wnab}\ups  + \gamma \left( \gamma' 
  + \frac{X I\alpha'}{\mu_0\sigma \lambda h} \right) \dd \ups \cdot\vec{\wnab}\ups 
  + \frac{\sigma n^2}{h} \Bigg\{ \frac{\lambda}{\gamma} 
  \bigg[ \Omega \Lambda' -  \Sigma' + \frac{I \alpha'}{\mu_0} 
  \nonumber \\
  & &  \qquad + \gamma' \lambda h (V - 2W \Omega - X \Omega^2) \bigg]  + T S' \Bigg\} 
  - \gamma \lambda n  \left( \Ci_{\vxi}
   + \Omega \Ci_{\vchi} \right) 
  + \frac{\gamma I^2}{\mu_0 \sigma \lambda h} \left( -X \Ci_{\vxi}
  + W \Ci_{\vchi} \right) = 0 ,   \label{e:master_transfield_tor}
\eea
whereas the poloidal wind equation (\ref{e:master_wind}) becomes
\be
  h^2 \left( \sigma + \frac{\gamma^2}{n^2} \, 
  \dd\ups \cdot\vec{\wnab}\ups \right)
  - \frac{1}{\gamma^2} \left( X\Sigma^2 + 2W \Sigma\Lambda
  - V\Lambda^2 \right)
  + \frac{I^2 h}{\mu_0 n} \left( 2 + \frac{X I^2}{\mu_0 \sigma^2 \lambda^2 n h } 
  \right) = 0 . 
\ee

\end{widetext}

\subsubsection{Pure rotational flow} \label{s:tor_rot}

In the particular case of a pure rotational flow ($\gamma = 0$), 
the transfield equation (\ref{e:master_transfield_tor}) reduces to 
[cf. Eq.~(\ref{e:transfield_rot}) with $\beta=0$]
\be \label{e:transfield_tor_rot_prov}
  I I' + \mu_0\sigma\left\{ (\vep + p) \left[ 
  \frac{K'}{K}
  + \lambda^2 (W+X\Omega)\Omega' \right] - n T S' \right\}= 0 , 
\ee
with $K(\ups)$ obeying to Eq.~(\ref{e:K_h}). 
In the present case, $\alpha = \beta = \gamma = 0$, i.e. 
$\dd\Phi = 0$, $\dd\Psi = 0$ and $\dd f = 0$. If the electromagnetic field is not vanishing, a natural choice for $\ups$ is
\be
  \ups = I . 
\ee
Then $I' = 1$ and Eq.~(\ref{e:transfield_tor_rot_prov}) reduces to 
\be \label{e:transfield_tor_rot}
  I  + \mu_0\sigma\left\{ (\vep + p) \left[ 
  \frac{K'}{K}
  + \lambda^2 (W+X\Omega)\Omega' \right] - n T S' \right\}= 0 . 
\ee
Given the functions $\Omega(I)$, $K(I)$ and $S(I)$, this equation has to be solved in 
$I$. Note that this is not a PDE in $I$ and 
that the matter quantities $n$, $\vep$, $p$ and $T$ are to be computed via the EOS from $S$ and $h$, the former being deduced from $K(I)$ and $\Omega(I)$
via Eq.~(\ref{e:K_h}): 
\be \label{e:h_K_Omega_I}
  h = \frac{K(I)}{\sqrt{V - 2W\Omega(I) - X\Omega(I)^2}} . 
\ee
In the special case $\Omega(I)=\mathrm{const}$ and $S(I)=\mathrm{const}$, we recover equations
obtained by Kiuchi and Yoshida (2008) \cite{KiuchY08} \footnote{The link between notations of 
Ref.~\cite{KiuchY08} and ours is $g_1 \leftrightarrow  -g/\sigma$, $g_2 \leftrightarrow  \sigma$, $\sqrt{g_2/g_1} F_{12} \leftrightarrow  I$, 
$u \leftrightarrow  \sigma n h = \sigma (\vep+p)$, $4\pi K(u)/u \leftrightarrow  -\mu_0 K'(I)/K(I)$.}.
At the Newtonian limit, the special case $\Omega(I)=\mathrm{const}$,  
$S(I)=\mathrm{const}$ and $K'(I)/K(I) = \mathrm{const}$ 
has been contemplated by Miketinac (1973) \cite{Miket73}.

It is worth underlining that in the case considered here, i.e. a pure rotational flow and a pure toroidal magnetic field, (i) the spacetime has to be circular (provided that 
the fluid and the electromagnetic field are the only sources in the Einstein equation) \cite{Oron02,KiuchY08}
and (ii) the twist functions $\Ci_{\vxi}$ and $\Ci_{\vchi}$, whose vanishing is equivalent
to circularity,  do not appear 
in Eqs.~(\ref{e:transfield_tor_rot})-(\ref{e:h_K_Omega_I}).

\subsection{Hydrodynamical limit} \label{s:hydro_lim}

\subsubsection{Generic case}

The hydrodynamical limit (no electromagnetic field) is easily taken by setting 
$\alpha = 0$ (i.e. $\dd\Phi=0$), $\beta = 0$ (i.e. $\dd\Psi=0$) and $I=0$ in
the equations obtained so far. 
In particular, the streamline-conserved quantities (\ref{e:Sigma_def}) and 
(\ref{e:Lambda_def}) reduce to 
\bea
  \Sigma & = & \gamma \, E \qquad\mbox{with}\qquad E = \lambda h(V - W\Omega) , \label{e:Sigma_hydro} \\
  \Lambda & = & \gamma \, L \qquad\mbox{with}\qquad L = \lambda h (W + X \Omega) , 
  \label{e:Lambda_hydro}
\eea
where the second equalities in each line follow from Eqs.~(\ref{e:E_express}) and
(\ref{e:L_express}) with $I=0$. 
Besides, Eq.~(\ref{e:def_A}) reduces to $A=\gamma^2$ and thanks to 
(\ref{e:Sigma_hydro})-(\ref{e:Lambda_hydro}), Eq.~(\ref{e:def_mA}) reduces to
$\mathcal{A} = \gamma (\Omega L' -E')$. 
Accordingly the master transfield equation (\ref{e:master_transfield}) becomes
\bea 
  & & \gamma^2 \Delta^* \ups + \gamma^2 \frac{n}{h} \, \dd\left(\frac{h}{n} \right) 
  \cdot \vec{\wnab} \ups + \gamma\gamma' \dd\ups \cdot\vec{\wnab} \ups \nonumber \\
  && \quad + \frac{\sigma n^2}{h} \left[ \lambda(\Omega L' -E') + T S' \right]
  - \gamma \lambda n  \left( \Ci_{\vxi}
   + \Omega \Ci_{\vchi} \right) = 0 . \nonumber \\
  & & \label{e:master_transfield_hydro}
\eea
On its side, the poloidal wind equation (\ref{e:master_wind}) reduces to
\be \label{e:master_wind_hydro}
  \frac{\gamma^2 h^2}{n^2} \, 
  \dd\ups \cdot\vec{\wnab}\ups + \sigma h^2 
  - X E^2 - 2W E L + V L^2 = 0 . 
\ee

\subsubsection{Flow with meridional component}

If the meridional fluid velocity is not vanishing, $\dd f \not = 0$ and a natural choice for the master potential is $\ups = f$. Then $\gamma = 1$ and 
Eqs.~(\ref{e:master_transfield_hydro}) and (\ref{e:master_wind_hydro}) become
\bea
  & & \Delta^* f +  \frac{n}{h} \, \dd\left(\frac{h}{n} \right) 
  \cdot \vec{\wnab} f 
  + \frac{\sigma n^2}{h} \left[ \lambda(\Omega L' -E') + T S' \right] \nonumber \\
  & & \qquad - \lambda n  \left( \Ci_{\vxi}
   + \Omega \Ci_{\vchi} \right) = 0 \label{e:transfield_hydro_f} \\
  & & \frac{h^2}{n^2} \, 
  \dd f \cdot\vec{\wnab} f + \sigma h^2 
  - X E^2 - 2W E L + V L^2 = 0 . \label{e:wind_hydro_f} 
\eea
These equations are to be supplemented by (i) Eq.~(\ref{e:lambda}) expressing $\lambda$ in terms
of $f$ and $\Omega$ [via Eq.~(\ref{e:w2_df2})]
and (ii) the hydrodynamical limit of Eq.~(\ref{e:Omega_M_A}), 
which reads
\be \label{e:Omega_hydro}
  \Omega = \frac{VL - WE}{XE + WL} .
\ee
It is then clear that, given the metric, the three functions $E(f)$, $L(f)$ and $S(f)$ and the
EOS $h=h(S,n)$, $T=T(S,n)$, 
Eqs.~(\ref{e:transfield_hydro_f})-(\ref{e:wind_hydro_f}) form a system of coupled PDE
for $(f,n)$. Solving this system provides a solution of the Euler equation for 
a rotating flow with meridional component. 
In the case of circular spacetimes ($\Ci_{\vxi} = \Ci_{\vchi} = 0$), Eq.~(\ref{e:transfield_hydro_f}) has been written first by Anderson (1989) \cite{Ander89}
and Beskin \& Pariev (1993) \cite{BeskiP93}, with $n$ extracted from 
Eq.~(\ref{e:wind_hydro_f}) and substituted in (\ref{e:transfield_hydro_f}), so that the system reduces to a single equation for $f$. 
An equivalent formulation has been developed recently by Birkl et al. \cite{BirklSM11} for
the barotropic case $(S=0)$, using the function $\psi := - \int E(f)\, df$ instead of $f$. 

In the Newtonian limit [cf. Eqs.~(\ref{e:transfield_Newt}) and (\ref{e:master_wind_Newt})],
the system becomes
\bea
    & & \Delta^* f - \frac{1}{n} \, \dd n  
  \cdot \vec{\wnab} f \nonumber \\
 & & \qquad  + \frac{n^2}{m_{\rm b}} \left[ \frac{L L'}{m_{\rm b}} 
  - r^2\sin^2\theta \left( E' - T S' \right)\right]  = 0  \label{e:transfield_hydro_f_Newt} \\
  & & \frac{1}{n^2} \, 
  \dd f \cdot\vec{\wnab} f + 2 r^2\sin^2\theta 
  \left( H + \Phi_{\rm grav} + 1 - \frac{E}{m_{\rm b}} \right) \nonumber \\
  & & \qquad 
  + \left( \frac{L}{m_{\rm b}} \right) ^2 = 0 \label{e:wind_hydro_f_Newt} .   
\eea
Note that we have substituted $\Omega$ by the Newtonian limit of (\ref{e:Omega_hydro}): 
$\Omega = L / (m_{\rm b} r^2 \sin^2\theta)$. 
Equation~(\ref{e:transfield_hydro_f_Newt}) is called \emph{Stokes equation}. 
For an incompressible fluid, it coincides with
Eq.~(7.5.11) of Batchelor treatise \cite{Batch70}. 
For a compressible fluid, we recover
Eq.~(1.107) of Beskin textbook \cite{Beski10} or Eq.~(15) of Eriguchi et al. \cite{EriguMH86}. 
Consequently, we shall call Eq.~(\ref{e:transfield_hydro_f}) the 
\emph{relativistic Stokes equation}. 

\subsubsection{Pure rotational flow}

For a pure rotational flow, $\gamma =0$ and Eqs.~(\ref{e:master_transfield_hydro}) and (\ref{e:master_wind_hydro}) reduce to 
\bea
  & & \Omega L' -E' + \frac{T}{\lambda} S' = 0 \label{e:transfield_hydro_rot} \\
  & & \sigma h^2 = X E^2 + 2 W E L - V L^2 . 
\eea
Let us assume that 
\be \label{e:hydro_Omega_ups}
  \Omega = \Omega(\ups) . 
\ee
In Sec.~\ref{s:rot_flow}, we have seen that this condition is mandatory if $\alpha\not=0$
or $\beta\not =0$. In the absence of electromagnetic field, (\ref{e:hydro_Omega_ups})
is fulfilled for rigid rotation ($\Omega=\mathrm{const}$) and for non-rigid one, it may be
imposed by a proper choice of $\ups$, for instance $\ups = \Omega$.
It is then natural to introduce, as in Sec.~\ref{s:rot_flow}, 
$K = K(\ups) = E - \Omega L = h / \lambda$ [cf. Eq.~(\ref{e:K_h})], 
so that Eq.~(\ref{e:transfield_hydro_rot}) can be written as
\be \label{e:Kp_Lom}
  K' + L \Omega' - \frac{T}{\lambda} S' = 0 . 
\ee
A wide class of equilibrium configurations is obtained by assuming 
\be \label{e:def_barT}
  \frac{T}{\lambda} = \bar T(\ups), 
\ee
with $\bar T$ is a function of $\ups$ such that 
\be \label{e:der_barT}
  {\bar T}' = - \bar T \frac{L}{K} \Omega' . 
\ee
Thanks to Eqs.~(\ref{e:def_barT})-(\ref{e:der_barT}), Eq.~(\ref{e:Kp_Lom}) can be written 
as
\be \label{e:Euler_hydro_rot}
   \left( \ln \bar\mu \right)' + \frac{L}{K} \Omega' = 0 . 
\ee
where 
\be
  \bar\mu = \bar\mu(\ups) := K - \bar T S = \frac{h - TS}{\lambda} = \frac{\mu}{\lambda}, 
\ee
$\mu$ being the baryon chemical potential introduced in Eq.~(\ref{e:def_T_mu}), the last equality
resulting from Eq.~(\ref{e:def_h}). 

If $\Omega$ is constant (rigid rotation), Eq.~(\ref{e:Euler_hydro_rot}) yields the first integral
\be \label{e:int_prem_rot_rigid}
  \bar \mu = \mathrm{const} . 
\ee
Note that in this case, Eq.~(\ref{e:der_barT}) implies the
so-called \emph{relativistic isothermal} condition : 
\be \label{e:isothermal}
  \bar T = \mathrm{const} . 
\ee

If $\Omega'\not=0$, Lemma~\ref{l:functions} of Sec.~\ref{s:sec_lemma} implies that 
$L/K$ must be a function of $\Omega$, $\mathcal{F}(\Omega)$, say. Since 
$L$ is expressible as (\ref{e:Lambda_hydro}) and $K=h/\lambda$
with the value (\ref{e:lambda_rot}) for $\lambda$, we have
\be
  \mathcal{F}(\Omega) = \frac{W+X\Omega}{V-2W\Omega - X\Omega^2} . 
\ee
Given the function $\mathcal{F}(\Omega)$, the above equation has to be solved in 
$\Omega$. At the Newtonian limit, it leads to a solution of the form 
$\Omega = \Omega(r\sin\theta)$, i.e. satisfying to the 
\emph{Poincar\'e-Wavre property} \cite{Tasso00}.  
With $L/K=\mathcal{F}(\Omega)$, Eq.~(\ref{e:Euler_hydro_rot}) is integrated to 
\be \label{e:int_prem_rot_diff}
  \ln\bar\mu + \int_0^{\Omega} \mathcal{F}(\tilde\Omega) \, d\tilde\Omega = 
  \mathrm{const}, 
\ee
and Eq.~(\ref{e:der_barT}) to 
\be
  {\bar T} \, \mathrm{e}^{\int_0^{\Omega} \mathcal{F}(\tilde\Omega) \, d\tilde\Omega }
  = \mathrm{const} ,
\ee
generalizing the isothermal condition (\ref{e:isothermal}) to the case of
differential rotation. 
In the case $T=0$ (or $S=\mathrm{const}$), we recognize in Eqs.~(\ref{e:int_prem_rot_rigid}) and (\ref{e:int_prem_rot_diff}) the standard first integrals governing rotating relativistic stars (see e.g. 
\cite{Sterg03} or \cite{Gourg10} and references
therein). For the finite temperature case, we recover results of 
Ref.~\cite{GoussHZ97-98}.


\section{Summary and conclusion} \label{s:concl}

We have formulated GRMHD for stationary and axisymmetric spacetimes in the most general case, i.e. non assuming circularity (as in Kerr spacetime). Moreover, we have based our approach on geometric quantities defined solely in terms of the spacetimes symmetries
(represented by the two Killing vectors $\vxi$ and $\vchi$), without relying on any coordinate system or any extra structure (such as a (2+1)+1 foliation). This provides some insight on previously introduced quantities and leads to the formulation of very general laws,
recovering previous ones as subcases and obtaining new ones in some specific limits. 
To our knowledge, the new results obtained here are: 
\begin{itemize}
\item the expression (\ref{e:F_gal}) of the electromagnetic field tensor $\w{F}$ entirely in terms of the two Killing vector fields and three scalar fields, independently of any coordinate system; 
\item the derivation of the conservation laws for the Bernoulli-type quantities $E$ and $L$ in a covariant manner and in the most general case, including that of a purely toroidal magnetic field disregarded in the original BO derivation \cite{BekenO78}; 
\item the fully covariant master transfield equation (\ref{e:master_transfield}) governing
the most general  
MHD equilibria in generic (i.e. noncircular) spacetimes, generalizing Soloviev 
non-relativistic equation \cite{Solov67}; 
\item the explicit form (\ref{e:gen_GS}) of the covariant Grad-Shafranov equation
for noncircular spacetimes;
\item the equation (\ref{e:master_transfield_tor}) governing MHD equilibria with
purely toroidal magnetic field in stationary and axisymmetric spacetimes;
\item the relativistic Stokes equation (\ref{e:transfield_hydro_f}) governing
hydrodynamical equilibria of flows with meridional components in stationary and axisymmetric spacetimes.
\end{itemize}
The relativistic master transfield equation (\ref{e:master_transfield}) is probably the
most important outcome of the present study. Beyond the aesthetic feature of having
a single equation governing all MHD equilibria, reducing to the relativistic 
Grad-Shafranov and Stokes equations in certain limits, 
the value of this equation resides in its potentiality to lead to solutions that
cannot be obtained by merely setting 
$\ups=\Psi$ or $\ups = f$, as already shown in the 
Newtonian regime \cite{GebhaK92}. 

In this article, we have focused on the derivation of the equations governing
MHD equilibria and of conservation laws. In order to solve the obtained equations, there
remains to choose the streamline-conserved functions
$\alpha(\ups)$, $\beta(\ups)$, $\gamma(\ups)$, $\Sigma(\ups)$, $\Lambda(\ups)$ and 
$S(\ups)$ and to specify some boundary conditions on them. An example of numerical resolution of the Grad-Shafranov equation (case $\ups = \Psi$) is provided by 
Ref.~\cite{IokaS04}. 
Finding stationary and axisymmetric GRMHD solutions provides initial data for 
dynamical stability studies of magnetized neutron stars (see e.g. \cite{KiuchSY08,KiuchSYS10,LieblLNP10,BucciD11}). 

As a final remark, let us point out that we have hardly used the axisymmetric character
of the Killing vector $\vchi$ (i.e. the fact that it is a generator of a
$\mathrm{SO}(2)$ group action), so that most results presented here would remain 
valid for any other type of spatial symmetry, like for instance translational symmetry.


\acknowledgments
We thank Silvano Bonazzola, Brandon Carter, John Friedman, Jean Heyvaerts, Kunihito Ioka and Christophe Sauty for very fruitful discussions. 
This work was supported by 
JSPS Grant-in-Aid for Scientific Research(C) 20540275, 
MEXT Grant-in-Aid for Scientific Research
on Innovative Area 20105004, 
NSF Grant PHY100155, the Greek State Scholarships Foundation
and ANR grant 06-2-134423 \emph{M\'ethodes
math\'ematiques pour la relativit\'e g\'en\'erale}.  
KU and EG acknowledge  support from a JSPS Invitation Fellowship 
for Research in Japan (Short-term) and the invitation program 
of foreign researchers at the Paris observatory.
 CM  thanks the Paris Observatory and the University of Wisconsin-Milwaukee for travel support.%


\appendix

\section{Lie derivative} \label{s:Lie_deriv}

The Lie derivative is the natural operator to express symmetries under a 1-parameter group action. It measures the change of a tensor field along the orbits of the group action. More precisely, any regular vector field $\vv{v}$ can be regarded as the generator of a 1-parameter group action on $\M$. Then, given a local coordinate system
$(x^\alpha)$ adapted to $\vv{v}$, i.e. such that $v^\alpha = (1,0,0,0)$, the Lie derivative along $\vv{v}$ of a tensor field $\w{T}$ of type $(k,\ell)$ is the tensor
field of the same type, whose components are the derivatives of $\w{T}$'s components  
with respect to the parameter associated with $\vv{v}$ (i.e. the coordinate $x^0$): 
\be
        (\Lie{\vv{v}} T)^{\alpha_1\ldots\alpha_k}_{\qquad\ \; \beta_1\ldots\beta_\ell}
        = \partial_0 T^{\alpha_1\ldots\alpha_k}_{\qquad\ \; \beta_1\ldots\beta_\ell} .
\ee
In an arbitrary coordinate system, this formula becomes
\bea
  & &   (\Lie{\vv{v}} T)^{\alpha_1\ldots\alpha_k}_{\qquad\ \; \beta_1\ldots\beta_\ell}
         =  v^\mu \partial_\mu T^{\alpha_1\ldots\alpha_k}_{\qquad\ \; \beta_1\ldots\beta_\ell} \nonumber \\
 & & \qquad \qquad - \sum_{i=1}^k T^{\alpha_1\ldots
\!{{{\scriptstyle i\atop\downarrow}\atop \scriptstyle\sigma}\atop\ }\!\!
\ldots\alpha_k}_{\qquad\ \ \ \  \  \  \beta_1\ldots\beta_\ell}
 \; \partial_\sigma v^{\alpha_i} \nonumber \\
        & &  \qquad \qquad +  \sum_{i=1}^\ell T^{\alpha_1\ldots\alpha_k}_{\qquad\ \; \beta_1\ldots
\!{\ \atop {\scriptstyle\sigma \atop {\uparrow\atop \scriptstyle i}} }\!\!
\ldots\beta_\ell} 
\; \partial_{\beta_i} v^{\sigma} .  \label{e:Lie_der_comp}
\eea 
In particular, for a scalar field $f$, 
\be
        \Lie{\vv{v}} f = v^\mu \partial_\mu f ,
\ee
for a vector field $\vv{w}$,
\be \label{e:Lie_der_vect}
        (\Lie{\vv{v}} \vv{w})^\alpha = v^\mu \partial_\mu w^\alpha
        - w^\mu \partial_\mu v^\alpha, 
\ee
for a 1-form $\vv{\omega}$, 
\be \label{e:Lie_der_1form}
        (\Lie{\vv{v}} \w{\omega})_\alpha = v^\mu \partial_\mu \omega_\alpha
        + \omega_\mu \partial_\alpha v^\mu , 
\ee
and 
for a bilinear form $\w{T}$ (such as the metric tensor $\w{g}$
or the electromagnetic field $\w{F}$), 
\be
        (\Lie{\vv{v}} \w{T})_{\alpha\beta} = v^\mu \partial_\mu T_{\alpha\beta}
        + T_{\mu\beta} \partial_\alpha v^\mu 
        + T_{\alpha\mu} \partial_\beta v^\mu . \label{e:Lie_der_bilin}
\ee
From formula (\ref{e:Lie_der_vect}), note that the Lie derivative of $\vv{w}$
along $\vv{v}$ is nothing but the commutator of the vector fields $\vv{v}$ and 
$\vv{w}$: 
\be \label{e:Lie_der_commut}
        \Lie{\vv{v}} \vv{w} = [\vv{v},\vv{w}] . 
\ee
Note also that in formulas (\ref{e:Lie_der_comp})-(\ref{e:Lie_der_bilin}), one may 
replace the partial derivative operator $\partial$ by the covariant derivative 
operator $\wnab$ associated with the metric $\w{g}$. This stems from the symmetry property of the Christoffel symbols. 
In particular, Eq.~(\ref{e:Lie_der_bilin}) can be written
\be
        (\Lie{\vv{v}} \w{T})_{\alpha\beta} = v^\mu \nabla_\mu T_{\alpha\beta}
        + T_{\mu\beta} \nabla_\alpha v^\mu 
        + T_{\alpha\mu} \nabla_\beta v^\mu . \label{e:Lie_der_bilin_nab}
\ee

\section{Differential forms and exterior calculus} \label{s:diff_geom}

Given a integer $p$ satisfying $0\leq p \leq 4$, a \emph{$p$-form} is a tensor of
type $(0,p)$ which is fully antisymmetric. By convention, 
a $0$-form is a scalar and a $1$-form is a linear form.
A \emph{differential form of rank $p$} is a field of $p$-forms over $\M$. 
Differential forms play a special role in the theory of integration on a 
manifold. Indeed the primary definition of an integral over a manifold of 
dimension $p$ is the integral of a $p$-form.
At a given point $x\in\M$, the set of all $1$-forms is $\T_x(\M)^*$,
the dual vector space to the vector space $\T_x(\M)$ tangent to $\M$ at $x$. 
More generally the set $\mathcal{A}^p_x(\M)$ of all $p$-forms at $x$ is a
vector space  of 
dimension $\binom{4}{p}$. In particular, the dimension of the space of 
4-forms is 1: all 4-forms are proportional to each other. 

We assume that the manifold $\M$ is \emph{orientable}, i.e. that 
there exists a continuous, nowhere vanishing, 4-form field. We may then introduce 
the \emph{Levi-Civita alternating tensor} $\weps$
(also called \emph{metric volume element}) as the differential form of rank $4$ such that
for any vector basis $(\vv{e}_\alpha)$ that is orthonormal with respect to $\w{g}$, 
\be \label{e:eps_bo}
  \weps(\vv{e}_0,\vv{e}_1,\vv{e}_2,\vv{e}_3) = \pm 1 . 
\ee
If $\M$ is orientable, there are actually two such 4-form fields, opposite to each other: picking one of them is making a choice of orientation on $\M$.
Having a $+$ sign (resp. $-$ sign) in Eq.~(\ref{e:eps_bo})
defines then a \emph{right-handed basis} (resp. a \emph{left-handed basis}). 
The components of $\weps$ in a given right-handed basis (not necessarily orthonormal) are
\be \label{e:eps_comp}
  \eps_{\alpha\beta\gamma\delta} = \sqrt{-g} [\alpha,\beta,\gamma,\delta] , 
\ee
where $g$ is the determinant of the components $(g_{\alpha\beta})$ of the metric tensor in 
the considered basis and $[\alpha,\beta,\gamma,\delta]$ stands for $1$ (resp. $-1$) if 
$(\alpha,\beta,\gamma,\delta)$ is an even (resp. odd) permutation of $(0,1,2,3)$, and 
$0$ otherwise. 

Two algebraic operations are defined on differential forms: the exterior product and 
Hodge star. The \emph{exterior product}  associates 
to any $p$-form $\w{A}$ and any $q$-form $\w{B}$  the $(p+q)$-form $\w{A}\wedge\w{B}$
defined by 
\begin{widetext}
\be \label{e:ext_prod}
        \w{A}\wedge\w{B}(\vv{v}_1,\ldots,\vv{v}_{p+q})
        := \frac{1}{p! q!} \sum_{\sigma\in\mathfrak{S}_{p+q}}
        (-1)^{k(\sigma)}  
        \w{A}(\vv{v}_{\sigma(1)},\ldots,\vv{v}_{\sigma(p)})
        \times \w{B}(\vv{v}_{\sigma(p+1)},\ldots,\vv{v}_{\sigma(p+q)})  ,
\ee
where $\vv{v}_1$,...,$\vv{v}_{p+q}$ are generic $p+q$ vectors,
$\mathfrak{S}_{p+q}$ is the group of permutations of $p+q$ elements, 
$(-1)^{k(\sigma)}$ is the signature of permutation $\sigma$ and 
and $\times$ denotes the 
multiplication in $\R$. 
In particular, if $\w{A}$ and $\w{B}$ are 1-forms, the exterior product is expressible in terms of tensor products as 
\be
  \w{A} \wedge \w{B} = \w{A}\otimes \w{B} - \w{B}\otimes \w{A} \quad \mbox{(1-forms)} . 
\ee
If $\w{A}$ is a 1-form and $\w{B}$ is a 2-form, then 
\be \label{e:ext_prod_1_2}
  \w{A}\wedge\w{B}(\vv{v}_1,\vv{v}_2,\vv{v}_3)
  = (\w{A}\cdot\vv{v}_1) \, \w{B}(\vv{v}_2,\vv{v}_3)
  + (\w{A}\cdot\vv{v}_2) \, \w{B}(\vv{v}_3,\vv{v}_1)
  + (\w{A}\cdot\vv{v}_3) \, \w{B}(\vv{v}_1,\vv{v}_2) .
\ee
\end{widetext}

The \emph{Hodge star} operator relies on the Levi-Civita tensor $\weps$: it associates
to every $p$-form $\w{\omega}$, a $(4-p)$-form $\star\w{\omega}$, called the 
\emph{Hodge dual of} $\w{\omega}$, and  defined by 
\bea
  \mbox{0-form} &:&  (\star\w{\omega})_{\alpha\beta\gamma\delta} =      
              \omega \, \eps_{\alpha\beta\gamma\delta}  \\
  \mbox{1-form} &:&  (\star\w{\omega})_{\alpha\beta\gamma} = \omega_\mu \, \epsilon^\mu_{\ \, \alpha\beta\gamma}   \label{e:Hodge_1} \\
  \mbox{2-form} &:&  (\star\w{\omega})_{\alpha\beta} = \frac{1}{2} \omega_{\mu\nu} \,
          \eps^{\mu\nu}_{\quad \alpha\beta} \label{e:Hodge_2}\\
  \mbox{3-form} &:&  (\star\w{\omega})_\alpha = \frac{1}{6} \omega_{\mu\nu\rho} \, \eps^{\mu\nu\rho}_{\quad\ \alpha}    \\
  \mbox{4-form} &:&  \star\w{\omega} =  \frac{1}{24} \omega_{\mu\nu\rho\sigma} \, \eps^{\mu\nu\rho\sigma} .
\eea
Notice that, for any $p$-form, 
\be \label{e:star_star}
  \star\star \omega = (-1)^{p+1} \omega 
\ee
and that, for any couple $(\w{a},\w{b})$ of 1-forms, 
\be \label{e:star_wedge}
  \star (\w{a}\wedge\w{b}) = \weps(\vw{a},\vw{b},.,.)
  \quad\mbox{and}\quad \star[\weps(\vw{a},\vw{b},.,.)] = - \w{a}\wedge\w{b},
\ee
where $\vw{a}$ (resp. $\vw{b}$) is the vector associated to $\w{a}$ (resp.
$\w{b}$) by the metric
[cf. Eq.~(\ref{e:def_vec_form})]. 

Being tensor fields, the differential forms are subject to the covariant derivative $\wnab$ and
to the Lie derivative $\Lie{\vv{v}}$ discussed above. 
But, in addition, they are subject to a third type
of derivation, called \emph{exterior derivation}.
The \emph{exterior derivative} of a $p$-form field $\w{\omega}$ is a
$(p+1)$-form field denoted $\dd\w{\omega}$. 
In terms of components with respect to a given
coordinate system $(x^\alpha)$, $\dd\w{\omega}$ is defined by
\bea
        \mbox{0-form} &:& (\dd\w{\omega})_\alpha = 
                \partial_\alpha\omega \label{e:def_ext_0f} \\
        \mbox{1-form}  &:&  (\dd\w{\omega})_{\alpha\beta} =
        \partial_\alpha\omega_\beta - \partial_\beta\omega_\alpha
                         \label{e:def_ext_1f} \\
        \mbox{2-form}  &:&  (\dd\w{\omega})_{\alpha\beta\gamma} =
        \partial_\alpha\omega_{\beta\gamma} + 
        \partial_\beta\omega_{\gamma\alpha} + 
        \partial_\gamma\omega_{\alpha\beta}  \label{e:def_ext_2f} \\
        \mbox{3-form}  &: & (\dd\w{\omega})_{\alpha\beta\gamma\delta} =
        \partial_\alpha\omega_{\beta\gamma\delta} 
        - \partial_\beta\omega_{\gamma\delta\alpha}  
         \nonumber \\
  & & \qquad\qquad\quad + \partial_\gamma\omega_{\delta\alpha\beta} - \partial_\delta\omega_{\alpha\beta\gamma} \label{e:def_ext_3f} . 
\eea
It can be easily checked that these formulas, although expressed in terms of 
partial derivatives of components in a coordinate system, do define tensor fields.
Notice that for a scalar field ($0$-form), the exterior derivative is nothing but the gradient 
$1$-form. Notice also that the definition of the exterior derivative appeals only to the
manifold structure. It does not depend upon the metric tensor  $\w{g}$, nor upon 
any other extra structure on $\M$.
Besides, as for the Lie derivative expressions (\ref{e:Lie_der_comp})-(\ref{e:Lie_der_bilin}), all partial derivatives in formulas 
(\ref{e:def_ext_0f})-(\ref{e:def_ext_3f}) can be replaced by covariant derivatives
$\nabla$ thanks to the symmetry of the Christoffel symbols.

A fundamental property of the exterior derivation is to be nilpotent:
\be \label{e:ext_der_nilpot}
         \dd\dd\w{\omega} = 0 .
\ee
A $p$-form $\w{\omega}$ is said to be \emph{closed} iff $\dd\w{\omega}=0$,
and \emph{exact} iff there exists a $(p-1)$-form $\w{\sigma}$ such that
$\w{\omega} = \dd\w{\sigma}$. From property (\ref{e:ext_der_nilpot}),
any exact $p$-form is closed. The \emph{Poincar\'e lemma} states that the converse is true,
at least locally. 

With respect to the exterior product, the exterior derivation obeys to a modified 
Leibniz rule: if $\w{a}$ is a $p$-form and $\w{b}$ a $q$-form, 
\be \label{e:Leibniz_wedge}
        \dd(\w{a}\wedge\w{b}) = \dd\w{a}\wedge \w{b} + (-1)^p \, \w{a}\wedge\dd\w{b} . 
\ee
If $p$ is even, we recover the standard Leibniz rule. 

The exterior derivative enters in the well known \emph{Stokes' theorem}: if $\mathcal{D}$
is a submanifold of $\M$ of dimension $d\in\{1,2,3,4\}$ and has a boundary (denoted $\partial\mathcal{D}$), then for any $(d-1)$-form $\w{\omega}$,
\be \label{e:Stokes}
        \oint_{\partial\mathcal{D}} \w{\omega} = 
        \int_{\mathcal{D}} \dd\w{\omega} .
\ee
Note that $\partial\mathcal{D}$ is a manifold of dimension $d-1$ and 
$\dd\w{\omega}$ is a $d$-form, so that each side of
(\ref{e:Stokes}) is a well defined quantity,
as the integral of a $p$-form over a $p$-dimensional manifold.

A standard identity relates the divergence of a vector field $\vv{v}$ 
to the exterior derivative of the 3-form $\vv{v}\cdot\weps$ 
(see e.g. Appendix~B of Ref.~\cite{Wald84}): 
\be  \label{e:divergence_eps}
        \dd(\vv{v}\cdot\weps) = (\diver\vv{v})\, \weps . 
\ee

Another very useful formula where the exterior derivative enters is
\emph{Cartan identity}, which states that the Lie derivative of a $p$-form 
$\w{\omega}$ ($p\geq 1$) along a vector field $\vv{v}$ is expressible as
\be \label{e:Cartan}
      \Lie{\vv{v}}\w{\omega} = \vv{v}\cdot\dd\w{\omega}
        + \dd(\vv{v}\cdot\w{\omega}) .
\ee
Notice that for a 1-form, Eq.~(\ref{e:Cartan}) is readily obtained 
by combining Eqs.~(\ref{e:Lie_der_1form}) and (\ref{e:def_ext_1f}). 

\section{Relation between the scalar fields $\Phi$ and $\Psi$ and the electromagnetic 4-potential} \label{s:potential}

The electromagnetic 4-potential $\w{A}$ is not an observable and may not necessarily obey the  symmetries  \eqref{e:sym_F} of the electromagnetic field $\w{F} = \dd\w{A}$. 
However, for each Killing vector, one can  find a gauge transformation  such that the 4-potential obeys the corresponding symmetry. We demonstrate this for stationarity and axisymmetry.

Stationarity implies that
 $\Lie{\vxi} \w{F} = \Lie{\vxi}\dd  \w{A}=\dd\Lie{\vxi}\w{A}=0$.
If $\M$ is simply connected, then the Poincar\'e lemma implies that there exists a single-valued scalar $\mu$ such that
 $\Lie{\vxi} \w{A} =\dd \mu$, 
 but this quantity will be nonzero in general. However, there exists a class of gauge transformations $\w{A'}=\w{A}+\dd \nu$ such that 
 $\Lie{\vxi} \w{A'} =\Lie{\vxi}(\w{A}+\dd \nu)=\dd(\mu+\Lie{\vxi}\nu)=0$ provided that the scalar $ \nu$ satisfies  
$\mu+\Lie{\vxi} \nu=\rm{const.}$ This differential equation  may be  integrated along the integral curves of the timelike Killing vector $\vxi$. This procedure    eliminates the time-dependent part of $\w{A}$.

Axisymmetry implies the relation
 $\Lie{\vchi} \w{F}= \Lie{\vchi}\dd  \w{A'}=\dd\Lie{\vchi}\w{A'}=0$ which, by virtue of the Poincar\'e lemma, implies the existence of a single-valued scalar $\mu'$ such that
 $\Lie{\vchi} \w{A'} =\dd \mu'$, 
 but  this quantity will again be nonzero in general. However,  there exists another class of \textit{time-independent} gauge transformations $\w{A''}=\w{A'}+\dd \nu'$ such that 
 $\Lie{\vchi} \w{A''} =\Lie{\vchi}(\w{A'}+\dd  \nu')=\dd(\mu'+\Lie{\vchi} \nu')=0$ with a time-independent scalar $ \nu'$ (obeying 
$\Lie{\vxi} \nu'=0$
by assumption) that satisfies the  equation 
$\mu'+\Lie{\vchi} \nu'=\rm{const.}$  This differential equation  can be integrated along the integral curves of the axial Killing vector $\vchi$. The resulting gauge transformation eliminates the non-axisymmetric part of $\w{A'}$ while maintaining its stationarity.

This permits one to work in a gauge class within which the 4-potential
$\w{A}$ is  stationary and axisymmetric. From the Cartan identity, 
$\vxi\cdot\dd\w{A}+\dd(\vxi\cdot \w{A})=\Lie{\vxi}  \w{A}=0$, one then has
$\vxi\cdot \w{F}=-\dd A_t$ and similarly $\vchi\cdot \w{F}=-\dd A_\varphi$. Comparing these two equations to \eqref{e:def_Phi} and \eqref{e:def_Psi}
allows one to identify $A_t$ with $\Phi$ and $A_\varphi$ with $\Psi$, up to some additive constant, thereby demonstrating Eq.~\eqref{e:At_Aphi}.

\section{Kerr-Newman electromagnetic field} \label{s:Kerr-Newman}

The Kerr-Newman solution describes a charged rotating black hole.
In Boyer-Lindquist coordinates $(t,r,\theta,\ph)$, 
its electromagnetic field is \cite{MisneTW73} 
\be
  \w{F} = \frac{\mu_0 Q}{4\pi(r^2 + a^2\cos^2\theta)^2}
    \left(  \w{P} \wedge \dd t 
 + \w{R} \wedge \dd\theta \right) , 
\ee
where $Q$ is the total electric charge, $a := J/M$ the reduced angular momentum
of the black hole, $\w{P} := (r^2-a^2\cos^2\theta) \, \dd r - a^2 r \sin2\theta \, \dd\theta$ and $\w{R} := a(a^2\cos^2\theta -r^2) \sin^2\theta \, \dd r
  + a r (r^2 + a^2) \sin 2\theta \, \dd\theta$. 
Since the Kerr-Newman spacetime is circular, 
$\wxis = \dd t$ and $\wchis = \dd \ph$ [cf. Eq.~(\ref{e:wxis_circular})].
The comparison with (\ref{e:F_gal}) leads to 
\bea
  & &\Phi =  - \frac{\mu_0 Q}{4\pi} \, \frac{r}{r^2 + a^2 \cos^2\theta}, \quad
        \Psi = \frac{\mu_0 Q}{4\pi} \, \frac{a r\sin^2\theta}{r^2 + a^2 \cos^2\theta}, \nonumber \\
  &&    I = 0 . 
\eea
At the non-rotating limit $(a=0)$, this reduces to Reissner-Nordstr\"om solution: 
\be
        \Phi = - \frac{\mu_0}{4\pi} \frac{Q}{r}, \qquad \Psi = 0, \qquad I = 0 .  
\ee

\section{Component expressions with respect to a coordinate system} \label{s:comp}

In coordinates $(x^\alpha) = (t,x^1,x^2,\ph)$ adapted to stationarity
and axisymmetry (cf. Sec.~\ref{s:sta_axi}), the components of the Killing vectors
are $\xi^\alpha = (1,0,0,0)$ and $\chi^\alpha = (0,0,0,1)$ 
[cf. Eq.~(\ref{e:adapted_coord})]. From Eq.~(\ref{e:eps_comp}), the 2-form 
$\weps(\vxi, \vchi, . ,.)$ is expressible as 
\be \label{e:weps_dx1dx2}
  \weps(\vxi, \vchi, . ,.) = \sqrt{-g} \; \dd x^1 \wedge \dd x^2 .
\ee

\subsection{General spacetimes}

The components $F_{\alpha\beta}$ of the electromagnetic field are given
by $\w{F} = F_{\alpha\beta} \, \dd x^\alpha\otimes \dd x^\beta  = F_{|\alpha\beta|} \, \dd x^\alpha\wedge \dd x^\beta$, where $|\alpha\beta|$ means that the summation is limited to $\alpha<\beta$. Accordingly, 
substituting Eqs.~(\ref{e:wxis_dt}) and  (\ref{e:weps_dx1dx2}) into Eq.~(\ref{e:F_gal}) 
leads to
\begin{subequations}
\label{e:comp_Fab}
\bea
  F_{0a} & = & - \partial_a \Phi, \qquad a\in\{1,2\} \label{e:F_0a} \\
  F_{3a} & = & - \partial_a \Psi, \qquad a\in\{1,2\} \label{e:F_3a} \\
  F_{03} & = & 0 \\
  F_{12} & = & \frac{1}{\sigma} \Big[ \partial_1 \Phi (-X \xi_2 + W \chi_2)
  - \partial_2 \Phi (-X \xi_1 + W \chi_1) \nonumber \\
  & & \qquad + \partial_1 \Psi ( W \xi_2 + V \chi_2) - \partial_2 \Psi ( W \xi_1 + V \chi_1) 
   \nonumber \\
   & & \qquad + I \sqrt{-g} \Big]
\eea
\end{subequations}

Regarding the fluid 4-velocity $\vv{u}$, 
the coefficients $\lambda$ and $\Omega$ in the decomposition (\ref{e:u_expand}) are given by $\lambda = \wxis \cdot \vv{u}$ and $\lambda \Omega =  \wchis \cdot \vv{u}$.  
Using (\ref{e:wxis_uxi_uchi}) and (\ref{e:def_V})-(\ref{e:def_sigma}), we get
\bea
         u^0 & = & \lambda + \frac{1}{\sigma} (X\xi_a - W\chi_a) u^a \label{e:u0_lambda} \\
         u^3 & = & \lambda\Omega - \frac{1}{\sigma} (W\xi_a + V\chi_a) u^a .
                                \label{e:u3_Omega} 
\eea
From (\ref{e:u_expand}) and the fact that $\xi^a = 0$ and $\chi^a=0$
($a\in\{1,2\}$), we get 
\be
        w^a = u^a . 
\ee
From (\ref{e:u_expand}) and (\ref{e:u0_lambda})-(\ref{e:u3_Omega}), the remaining components
of $\vv{w}$ are
\bea
        w^0 & = & \frac{1}{\sigma} (X\xi_a - W\chi_a) u^a  \label{e:w0} \\
         w^3 & = & - \frac{1}{\sigma} (W\xi_a + V\chi_a) u^a .\label{e:w3} 
\eea

From the property $\vv{w}\in\Pi^\perp$, we get $w_0=\w_\mu \xi^\mu= 0$ and
$w_3 = w_\mu \chi^\mu = 0$, hence the covariant components of $\vv{w}$: 
\be
        w_\alpha = \left(0,\  u_a - \lambda(\xi_a + \Omega \chi_a),\ 0 \right) . 
\ee
The covariant components of $\vv{u}$ are given by 
$u_0 = u_\mu \xi^\mu$ and $u_3 = u_\mu \chi^\mu$, so that we may write
\be
        u_\alpha = \left(-\lambda(V-W \Omega),\, u_a,\, \lambda(W+X \Omega) \right) . 
\ee

\subsection{Circular spacetimes}

In the circular case, we may choose coordinates $(t,x^1,x^2,\ph)$ so that the surfaces orthogonal to $\Pi$ are the surfaces $\{t=\mathrm{const},\; \ph=\mathrm{const}\}$. Then
(\ref{e:gab_circular}) holds and we have $\xi_a = g_{a\mu} \xi^\mu = g_{a0} = 0$
and $\chi_a = g_{a\mu} \chi^\mu = g_{a3} = 0$:
\be
        \xi_a = \chi_a = 0 . 
\ee
Accordingly, the components (\ref{e:comp_Fab}) of the electromagnetic field simplify to 
\be
    F_{\alpha\beta} = \begin{pmatrix}
  0 & - \partial_1 \Phi & - \partial_2 \Phi & 0 \\
  \partial_1 \Phi & 0 & \sqrt{-g} I / \sigma & \partial_1 \Psi \\
  \partial_2 \Phi & - \sqrt{-g} I / \sigma & 0 & \partial_2 \Psi \\
  0 & -\partial_1 \Psi & - \partial_2 \Psi & 0 
  \end{pmatrix}
\ee
and relations (\ref{e:u0_lambda})-(\ref{e:u3_Omega}) reduce to 
\be
        u^0 = \lambda \qquad\mbox{and}\qquad u^3 = \lambda\Omega , 
\ee
whereas (\ref{e:w0})-(\ref{e:w3}) become
\be \label{e:w0_w3_null}
        w^0 = w^3 = 0 . 
\ee

\end{document}